\documentstyle[12pt]{article}
\begin{document}
\input vatola.sty
\renewcommand{\thesubsection}{\arabic{subsection}}
\renewcommand{\thesubsubsection}
             {\arabic{subsection}.\arabic{subsubsection}}
\renewcommand{\theequation}{\arabic{section}.\arabic{equation}}

\baselineskip=20pt

\thispagestyle{empty}

\vspace{10mm}

{\sf
\title{
{\normalsize
\begin{flushright}
CU-TP-1008\\
RBRC-176
\end{flushright}}
A Convergent Iterative Solution of the Quantum Double-well Potential\thanks{This
research was supported in part by U. S. Department of Energy grand \#DE-FG02-92ER40699,
and by the RIKEN-BNL Research Center, Brookhaven National Laboratory}}

\author{
R. Friedberg$^{1}$, T. D. Lee$^{1-3}$, W. Q. Zhao$^{2,4}$ and A. Cimenser$^{1}$\\
\\
{\small \it 1. Physics Department, Columbia University}\\
{\small \it          New York, NY 10027, USA}\\
{\small \it 2. China Center of Advanced Science and Technology (CCAST)}\\
{\small \it         (World Lab.), P.O. Box 8730, Beijing 100080,  China}\\
{\small \it 3. RIKEN BNL Research Center (RBRC), Brookhaven National Lab.}\\
{\small \it         Bldg. 510, BNL, Upton, NY 11943, USA}\\
{\small \it 4. Institute of High Energy Physics, Chinese Academy of Sciences} \\
{\small \it          P.O. Box 918(4), Beijing 100039, China}
  }
\maketitle

\newpage

\begin{abstract}
We present a new convergent iterative solution for the two lowest quantum wave
functions $\psi_{ev}$ and  $\psi_{od}$ of the Hamiltonian with a quartic double well
potential $V$ in one dimension. By starting from a trial function, which is by itself
the exact lowest even or odd eigenstate of a different Hamiltonian with a modified
potential $V+\delta V$, we construct the Green's function for the modified
potential. The true wave functions, $\psi_{ev}$ or $\psi_{od}$, then satisfies a
linear inhomogeneous integral equation, in which the inhomogeneous term is the trial
function, and the kernel is the product of the Green's function times the sum of
$\delta V$, the potential difference, and the corresponding energy shift. By
iterating this equation we obtain successive approximations to the true wave
function; furthermore, the approximate energy shift is also adjusted at each
iteration so that the approximate wave function is well behaved everywhere. We are
able to prove that this iterative procedure converges for both the energy and the
wave function at all $x$.

The effectiveness of this iterative process clearly depends on how good the trial
function is, or equivalently, how small the potential difference $\delta V$ is.
Although each iteration brings a correction smaller than the previous one by a
factor proportional to the parameter that characterizes the smallness of $\delta V$,
it is not a power series expansion in the parameter. The exact tunneling information
of the modified potential is, of course, contained in the Green's function; by
adjusting the kernel of the integral equation via the energy shift at each
iteration, we bring enough of this information into the calculation so that each
approximate wave function is exponentially tuned. This is the
underlying reason why the present method converges, while the usual power series
expansion does not.

\end{abstract}

\vspace{1cm}

\noindent
~~~~~~~~PACS{:~~03.65.Ge, 11.15.Me}

\newpage

\section*{\bf 1. Introduction}
\setcounter{section}{1}
\setcounter{equation}{0}

The quartic potential in one dimension with degenerate minima,
\begin{eqnarray}\label{e0.1}
V(x) = \frac{1}{2}g^2 (x^2-a^2)^2
\end{eqnarray}
is the archetypical bound-state tunneling problem in quantum mechanics. Its
perturbation series is well known but does not converge as one would expect since the
exponential contribution from tunneling between the two minima is necessarily
nonperturbative.

An alternative form of the same problem is obtained by setting $q=\sqrt{2ga}(a-x)$
so that the Hamiltonian becomes
\begin{eqnarray}\label{e0.2}
H=-\frac{1}{2}~\frac{d^2}{dx^2}+ V(x) \equiv 2gah,
\end{eqnarray}
where
\begin{eqnarray}\label{e0.3}
h=-\frac{1}{2}~\frac{d^2}{dq^2} + \frac{1}{2} q^2 \Big(1-\frac{q}{\sqrt{8ga^3}}\Big)^2.
\end{eqnarray}

This shows that the dimensionless (small) expansion parameter is related to
$1/\sqrt{8ga^3}$; as it turns out, the relevant parameter is its square. In this
paper we shall take $a=1$ so that the expansion parameter is $1/g$; in the
literature one often finds the assumption $2ga=1$ (placing the second minimum of the
potential at $q=1/g$) so that $1/\sqrt{8ga^3}$ reduces to $g$ and the anharmonic
potential appears as $(1/2)q^2(1-gq)^2$. Then $g$ appears with positive
exponents instead of negative, but the coefficients of the power series are the same
as with our form of the potential, apart from the overall factor $2ga$.

Recent work[1-10] on this problem has concentrated on the Feynman path integral by which
the nonperturbative tunneling effect is represented by a series of instantons and
anti-instantons. Emphasis is sometimes on understanding the asymptotic behavior of
the perturbation series as a signature of the tunneling influence, and sometimes on
supplying a nonperturbative correction to supplement the truncated perturbation
series.

In particular, Zinn-Justin[4-7] has produced an elegant implicit equation that determines
all the eigenenergies in terms of two functions given as power series: one series is
the Brillouin-Wigner expansion for the energy, and the other is the steepest-descent
expansion for the action of a single instanton. To any order in the power series, an
expression for the energy is obtained which, through the implicit equation, contains
information pertaining to {\it arbitrary} numbers of instantons. One is thus in the
odd position of possessing approximations
to finite order in $f=1/(ga^3)$ but in some sense correct to all orders
in $e^{-1/f}$.

We offer here a completely different method which leads to a result of similar
character although its equivalence with that of Zinn-Justin is yet to be studied.
Our method is based entirely on the Schroedinger equation, with no reference to path
integrals or to analytic continuation in complex $g$. We start from an approximation
to the desired wavefunction, which itself satisfies exactly a Schroedinger equation
with a modified potential $V+\delta V$. From this trial wavefunction
we obtain the Green's
function for the modified potential. The true wavefunction is then given implicitly
by acting with the Green's function on a source obtained by multiplying the
difference between the two potentials by the true wavefunction itself. By iterating
this equation we obtain successive approximations to the true wavefunction. We are
able to prove that these approximations actually converge.

The effectiveness of this iterative procedure clearly depends on how good
the trial function is, or equivalently, how small the potential difference $\delta V$
between the modified potential and the true potential is.
Although each iteration brings a correction smaller than the previous one by a
factor proportional to the parameter that characterizes the smallness of $\delta V$,
it is not a power series expansion of the parameter, since the Green's function also
contains the exact tunneling information of the modified potential. This is the
reason why the present method converges, while the usual power series
expansion does not.

Another important feature of the method is that the approximate energy is also adjusted
at each stage so as to prevent the approximate wavefunction from getting too large at
the second minimum. Thus we simultaneously develop a convergent sequence of
approximations for the energy. Like that of Zinn-Justin, the sequence is correct to
successively higher orders in $f$ (or, $1/g$ in our notation), but each step is in
some sense correct to all orders in $e^{-1/f}$.

Because we are directly dealing with the Schroedinger equation
\begin{eqnarray}\label{e0.4}
H\psi(x)=E \psi(x),
\end{eqnarray}
the proof of convergence is explicit, and it follows the traditional method of
analysis, as we shall see.

Recall that the formal perturbative expansion of
\begin{eqnarray}\label{e0.5}
\psi(x) = e^{-gS(x)}
\end{eqnarray}
and its associated energy $E$ may be written as
\begin{eqnarray}\label{e0.6}
gS=gS_0+S_1+\frac{1}{g}S_2+\cdots
\end{eqnarray}
and
\begin{eqnarray}\label{e0.7}
E=gE_0+E_1+\frac{1}{g}E_2+\cdots,
\end{eqnarray}
where (after setting the parameter $a=1$ in the potential $V(x)$)
\begin{eqnarray}\label{e0.8}
S_0(x) = \frac{1}{3}(x-1)^2(x+2),~~~~~S_1(x) = \ln \frac{x+1}{2},~~~~~
S_2(x) = \frac{3}{16}-\frac{x+2}{4(x+1)^2},~~~\cdots
\end{eqnarray}
and
\begin{eqnarray}\label{e0.9}
E_0=1,~~~E_1=-\frac{1}{4},~~~E_2=-\frac{9}{64},~~~\cdots.
\end{eqnarray}
Both expansions (\ref{e0.6}) and (\ref{e0.7}) are divergent; furthermore, at $x=-1$
and for $n \ge 1$, each $S_n(x)$ is infinite. The reflection $x \rightarrow -x$
gives a corresponding asymptotic expansion $S_n(x) \rightarrow S_n(-x)$, in which
each $S_n(-x)$ is regular at $x=-1$, but singular at $x=+1$.

In this paper, we will not use the perturbative expansion. Instead, we will
follow a new approach suggested in Section 3 and 4 of a previous paper[11]
( referred to as {\rm I} hereafter ). Accordingly, we will construct
an appropriate trial function to start a new iterative procedure to derive the
lowest even, or odd, state which satisfies
\begin{eqnarray*}
H\psi_{ev}=E_{ev} \psi_{ev}~~~~~~{\sf or}~~~~~~H\psi_{od}=E_{od}\psi_{od}~.
\end{eqnarray*}
As first step, we retain the first two terms in the expansion (\ref{e0.6}) of $S$.
For $x\geq 0$, the function
\begin{eqnarray*}
\phi_+(x)=e^{-gS_0(x)-S_1(x)}
\end{eqnarray*}
is well behaved and, for $g$ large, approximates the overall behavior of
$\psi_{ev}$. However, $\phi_+$ does not satisfy the boundary condition of
$\psi_{ev}$ at $x=0$.

  In Section 2, we use $\phi_+$ to construct a trial function $\phi$:\\
for $x\geq 0$
\begin{eqnarray*}
\phi(x)\equiv \left\{
\begin{array}{ll}
\phi_+(x) + \frac{g-1}{g+1}\phi_-(x), &~~~~ {\sf if} ~~~0\leq x<1\\
(1+ \frac{g-1}{g+1}e^{-\frac{4}{3}g})\phi_+(x), &~~~~ {\sf if} ~~~x>1
\end{array}
\right.
\end{eqnarray*}
where
\begin{eqnarray*}
\phi_-(x)\equiv
e^{-gS_0(-x)-S_1(x)}=e^{-\frac{4}{3}g+gS_0(x)}\frac{2}{1+x},
\end{eqnarray*}
for $x<0$
\begin{eqnarray*}
\phi(x)=\phi(-x).
\end{eqnarray*}
Thus, $\phi(x)$ is even in $x$ and at $x=0$, its derivative $\phi'(0)=0$.
Furthermore, it can be verified that $\phi(x)$ and its first derivative $\phi'(x)$
are continuous at all $x$, and in addition, $\phi(x)$ is the lowest eigenstate of a
Schroedinger equation with a different Hamiltonian:
\begin{eqnarray*}
\Big[-\frac{1}{2}\frac{d^2}{dx^2}+V(x)+ w(x)\Big]\phi(x)=g\phi(x)
\end{eqnarray*}
where except for $x$ near $\pm 1$,
\begin{eqnarray*}
\frac{w(x)}{V(x)}=O(\frac{1}{g})~~~~{\sf or~~smaller}.
\end{eqnarray*}
When $x$ is near $\pm 1$, $\phi(x)$ has the correct Gaussian behavior, like
$\psi_{ev}(x)$.

  Next, we construct the Green's function $(x|D|y)$, whose matrix form is defined by
\begin{eqnarray*}
D \equiv -2\phi\theta\phi^{-2}\theta\phi,
\end{eqnarray*}
in which the matrix element of $\theta$ is a step function:
\begin{eqnarray*}
(x|\theta|y) \equiv \left\{
\begin{array}{ll}
0, & ~~~~{\sf for}~~~ x>y\\
-1, & ~~~~ {\sf for}~~~ x<y.
\end{array}
\right.
\end{eqnarray*}
Thus, $D$ satisfies
\begin{eqnarray*}
\Big[-\frac{1}{2}\frac{d^2}{dx^2}+V(x)+ w(x)-g\Big](x|D|y)=\delta (x-y)
\end{eqnarray*}
and, in the matrix notation,
\begin{eqnarray*}
\psi_{ev} = \phi + D(w+E_{ev}-g)\psi_{ev}.
\end{eqnarray*}
The iterative procedure beginning from $\phi$ to arriving at $\psi_{ev}$ is discussed
in Section 2. The convergence of this procedure is proved in Section 3.

To demonstrate the flexibility of our approach, we give a variation of the procedure
for the derivation of the lowest odd eigenfunction $\psi_{od}$ of the double-well
potential problem. In Section 4 we begin with the trial function $\phi_+$. As we
shall see, using $\phi_+$, instead of $\phi$, as the zero$^{{\sf th}}$ approximation,
we can arrive at a solution $\psi_+(x)$ of the differential equation
\begin{eqnarray*}
H\psi_+(x) = \Big[-\frac{1}{2}\frac{d^2}{dx^2}+V(x)\Big]\psi_+(x)=E_+\psi_+(x)
\end{eqnarray*}
with $\psi_+(x)$ and its derivative $\psi'_+(x)$ satisfying the boundary conditions:
$$
{\sf at}~~~x=0,~~~~~\frac{\psi_+'(0)}{\psi_+(0)}=\frac{\phi_+'(0)}{\phi_+(0)}\\
$$
and
$$
{\sf at}~~~x=\infty,~~~~~\psi_+(\infty) = 0.
$$
(At $x=-\infty$, $\psi_+(x)$ diverges.) For $ x \ge 0$, the derivation of
$\psi_+$ from $\phi_+$
follows the same iterative steps as those in Sections 2 and 3, through the
replacement of $\phi$ by $\phi_+$ and $\psi_{ev}$ by $\psi_+$. Likewise, the
iteration is a convergent one.

In Section 5, we start from $\psi_+$ to first construct an odd trial function
$\chi$. The function $\chi$ is defined to be:\\
for $x\geq 0$
\begin{eqnarray*}
\chi=\psi_+ - \psi_-,
\end{eqnarray*}
where (in the matrix form)
\begin{eqnarray*}
\psi_- = 2 \gamma \psi_+\theta\psi_+^{-2}\theta\psi_+^2
\end{eqnarray*}
with the constant $\gamma$ determined by the condition: at $x=0$, $\chi(0) =0$; \\
for $x<0$
\begin{eqnarray*}
\chi(x) \equiv -\chi(-x).
\end{eqnarray*}
It will be shown that $\chi(x)$ is the lowest eigenstate of a new Hamiltonian:
\begin{eqnarray}\label{e0.10}
(H+\nu)\chi=E_+\chi,
\end{eqnarray}
where $H$ is the same double-well Hamiltonian, as before. As we shall see, $\nu$ is
$O(\epsilon)$, where $\epsilon=e^{-\frac{4}{3}g}$ is the very small tunneling
parameter, and $E_+$ is between the two eigenvalues $E_{ev}$ and $E_{od}$ of $H$.

Using $\chi$ as the new zero$^{{\sf th}}$ order approximation for $\psi_{od}$ we
first construct the corresponding Green's function $G$, defined by
\begin{eqnarray*}
G=-2\chi\theta\chi^{-2}\theta\chi.
\end{eqnarray*}
Because of (\ref{e0.10}), $G$ and $\psi_{od}$ satisfy
\begin{eqnarray*}
(H+\nu-E_+)G=1
\end{eqnarray*}
and
\begin{eqnarray}\label{e0.11}
\psi_{od} = \chi +G(\nu+E_{od}-E_+)\psi_{od}.
\end{eqnarray}

  The integral equation (\ref{e0.11}) now serves as the basis of a new iterative
procedure using $G(\nu+E_{od}-E_+)$ as the kernel for iterations. As will be shown,
each iteration improves the solution towards $\psi_{od}$ by a factor $O(\epsilon)$.
It will be proved that this new iterative sequence is also a convergent one.
Furthermore, the convergence is point-wise in all $x$ ( as are the previous iterations,
from $\phi$ to $\psi_{ev}$ and from $\phi_+$ to $\psi_+$).

  Throughout the paper, most
of the iterative functions are only known through their functional relations. To
achieve the convergence proof, we rely on deriving the explicit upper and lower
bounds for a number of these functions and their related parameters. Some of their
derivations can be rather lengthy; these are given in Appendices A to C.

From a practical point of view, once a good trial function is constructed, the new
iterative procedure is quite effective in generating better and better
approximations, as will be discussed in a particular example given in Appendix D.

\newpage

\section*{\bf 2. The Iterative Process}
\setcounter{section}{2}
\setcounter{equation}{0}

  With $a=1$, the one-dimensional Hamiltonian (\ref{e0.2}) becomes
\begin{eqnarray}\label{e1.1}
H=T+ \frac{1}{2}g^2 (x^2-1)^2,
\end{eqnarray}
where $T=-\frac{1}{2} \frac{d^2}{dx^2}$. Let $\psi_{ev}$ be the lowest eigen-state,
which is even in $x$ and satisfies
\begin{eqnarray}\label{e1.2}
H\psi_{ev}=E_{ev} \psi_{ev}.
\end{eqnarray}
As mentioned in the introduction, we shall solve $\psi_{ev}$ by starting
from a trial function $\phi(x)$ which satisfies a Schroedinger
equation with a different Hamiltonian say, instead of (\ref{e1.2}),
\begin{eqnarray}\label{e1.3}
(H+w)\phi=(E_{ev}+{\cal E}) \phi.
\end{eqnarray}
The function $\phi(x)$, like $\psi_{ev}(x)$, is even in $x$ and goes to $0$ at
$x=\pm  \infty$.
From $\phi(x)$, one can readily construct an irregular solution $\overline{\phi}(x)$
which satisfies the same equation (\ref{e1.3}):
\begin{eqnarray*}
\overline{\phi}(x)=\phi(x) \int\limits^x\phi^{-2}(y)dy
\end{eqnarray*}
The standard Green's function $(x|D|y)$ for the Sturm-Liouville type
problem[13] can be formed by using the product of the bilinear expression
\begin{eqnarray}
\phi(x)\overline{\phi}(y)-\overline{\phi}(x)\phi(y)\nonumber
\end{eqnarray}
times a step-function $\theta$ in $x-y$. As will be  discussed,  the
eigen-function $\psi_{ev}$ of   the original problem (\ref{e1.1})-
(\ref{e1.2}) can then
be derived as a power series of $D$. The success of the method clearly depends on the
choice of the trial function $\phi(x)$. The difference $w$ between these two Hamiltonian
in (\ref{e1.2}) and (\ref{e1.3}) can be obtained by a straightforward
differentiation of $\phi$, as we shall see.

  Following the steps outlined in the Introduction, we introduce for $x \geq 0$,
\begin{eqnarray}\label{e1.4}
g S_0(x) & \equiv & \frac{g}{3}(x-1)^2(x+2),\\
S_1(x) &\equiv & ln~\frac{x+1}{2},\\
\phi_+(x) &\equiv &  e^{-g S_0(x)-S_1(x)} =  e^{-g S_0(x)}(\frac{2}{1+x}),\\
\phi_-(x) &\equiv &  e^{-gS_0(-x)-S_1(x)}=
e^{-\frac{4}{3}g}e^{+gS_0(x)}(\frac{2}{1+x})
\end{eqnarray}
and
\begin{eqnarray}\label{e1.8}
\phi(x) = \phi(-x) \equiv
\left\{\begin{array}{ccc}
\phi_+(x) + \frac{g-1}{g+1} \phi_-(x),~~~~~~~{\sf for}~~0 \leq x <1\\
(1+\frac{g-1}{g+1}e^{-\frac{4}{3}g})\phi_+(x),~~~~~~{\sf for}~~x>1.
\end{array}
\right.
\end{eqnarray}
The functions $gS_0(x)$ and $S_1(x)$ are identical to those used in {\rm I} and Ref.[12]
(referred to as {\rm II} hereafter; note that the $\phi_+$ and $\phi_-$ in this paper
are different from those defined in {\rm II}.). By differentiating $\phi_+$, we find
that it satisfies
\begin{eqnarray}\label{e1.9}
(T+V+u)\phi_+=g\phi_+,
\end{eqnarray}
where
\begin{eqnarray}\label{e1.10}
V(x)= \frac{1}{2}g^2 (x^2-1)^2
\end{eqnarray}
and
\begin{eqnarray}\label{e1.11}
u(x)=\frac{1}{(1+x)^2}.
\end{eqnarray}

  A good trial function $\phi(x)$ for $\psi_{ev}(x)$ must also be even in $x$.
Thus, in the upper expression in
(\ref{e1.8}) the additional term proportional to $\phi_-(x)$ for $0 \leq x <1$ is
to make the derivative of $\phi(x)$ vanish at $x=0$; i.e.,
\begin{eqnarray}\label{e1.12}
\phi'(0)=0.
\end{eqnarray}
(Throughout the paper, prime denotes $\frac{d}{dx}$, and $g$ is $>1$.) At $x=0$,
$\phi_+(0)=\phi_-(0)=e^{-\frac{2}{3}g}$. As $x$ increases from $0$, $ \phi_+(x)$
increases to $1$ at $x=1$, while $ \phi_-(x)$ decreases to $e^{-\frac{4}{3}g}$
at $x=1$. We modify
$\phi(x)$ to the lower expression in (\ref{e1.8}) so that $\phi(x)$
becomes proportional to $\phi_+(x)$
for $x>1$; furthermore, $\phi(x)$ and $\phi'(x)$ are both continuous at $x=1$.

The function $\phi$ satisfies
\begin{eqnarray}\label{e1.13}
(T+V+w)\phi=g \phi,
\end{eqnarray}
where, for $x \geq 0$,
\begin{eqnarray}\label{e1.14}
w(x)=u(x) + \hat{g} (x)
\end{eqnarray}
with $u(x)$ and $\hat{g} (x)$ both even in $x$. For $x>0$, $u(x)$ is given by
(\ref{e1.11}) and
\begin{eqnarray}\label{e1.15}
\hat{g} (x) =
\left\{\begin{array}{ll}
2g\frac{(g-1) e^{2g S_0(x)-\frac{4}{3} g}}{(g+1)+(g-1)e^{2g S_0(x)-\frac{4}{3} g}},
&~~~~~~{\sf for}~~0 \leq x<1\\
0&~~~~~~{\sf for}~~x>1,
\end{array}
\right.
\end{eqnarray}
which is discontinuous at $x=1$. For $x<0$, $w(x)=w(-x)$.

Introduce the matrix $\theta$ whose matrix elements are
\begin{eqnarray}\label{e1.16}
(x|\theta|y)=
\left\{\begin{array}{ccc}
0~~~~~&{\sf for}~~~~~&x>y,\\
-1~~~~~&{\sf for}~~~~~&x<y.
\end{array}
\right.
\end{eqnarray}
Following the Green's function notations in I, we define the matrices
\begin{eqnarray}\label{e1.17}
\overline{D}~ \equiv -2 \theta \phi^{-2}\theta\phi^2
\end{eqnarray}
and
\begin{eqnarray}\label{e1.18}
D = \phi \overline{D}~ \phi^{-1}.
\end{eqnarray}
Differentiating $D$ with respect to  $x$ from the left, we find
\begin{eqnarray}
D' = -2\phi' \theta \phi^{-2} \theta \phi +2 \phi^{-1} \theta \phi \nonumber
\end{eqnarray}
and
\begin{eqnarray}
D'' = -2\phi'' \theta \phi^{-2} \theta \phi - 2. \nonumber
\end{eqnarray}
Thus,
\begin{eqnarray}\label{e1.19}
(T+V+w-g)D= I = {\sf unit~matrix},
\end{eqnarray}
and we see that $\psi_{ev}$ of (\ref{e1.2}) satisfies
\begin{eqnarray}\label{e1.20}
\psi_{ev} = \phi + D(w - {\cal E}) \psi_{ev},
\end{eqnarray}
with its eigenvalue $E_{ev}$ given by
\begin{eqnarray}\label{e1.21}
E_{ev} = g - {\cal E}.
\end{eqnarray}
In addition,
\begin{eqnarray}\label{e1.22}
{\cal E} = \frac{\int\limits_0^{\infty} w \phi \psi_{ev} dx}
{\int\limits_0^{\infty} \phi \psi_{ev} dx}.
\end{eqnarray}
The proof is identical to that given for (4.92) of I. In the following, we give an
alternative proof by introducing $f$ through
\begin{eqnarray}\label{e1.23}
\psi_{ev} = \phi f.
\end{eqnarray}
Thus, $f$ satisfies
\begin{eqnarray}\label{e1.24}
f = 1+ \overline{D} (w - {\cal E}) f;
\end{eqnarray}
i.e., on account of (\ref{e1.16}) and (\ref{e1.17}),
\begin{eqnarray}\label{e1.25}
f(x) = 1 -2\int\limits_x^{\infty} \phi^{-2}(y) dy \int\limits_y^{\infty} \phi^2(z)
(w(z) - {\cal E}) f(z)dz.
\end{eqnarray}
Since at $x=0$,  $\psi'_{ev}(0) = 0$ and $\phi'(0)=0$, we also have $f'(0)=0$,
which leads from (\ref{e1.25}) to
\begin{eqnarray}
\int\limits_0^{\infty} \phi^2(x)(w(x)-{\cal E})f(x) dx = 0,\nonumber
\end{eqnarray}
and gives (\ref{e1.22}).

In what follows, we shall solve $f$ and ${\cal E}$ by the iterative sequences
$\{f_n\}$ and $\{{\cal E}_n\}$:
\begin{eqnarray}
f_0 &=& 1,~f_1,~f_2,~f_3,~\cdots,~f_n,~\cdots \nonumber
\end{eqnarray}
and
\begin{eqnarray}\label{e1.26}
{\cal E}_0 &=& 0,~{\cal E}_1,~{\cal E}_2,~{\cal E}_3,~\cdots,~{\cal E}_n,~\cdots
\end{eqnarray}
We require
\begin{eqnarray}\label{e1.27}
f_n = 1 + \overline{D} (w-{\cal E}_n) f_{n-1}
\end{eqnarray}
with
\begin{eqnarray}\label{e1.28}
{\cal E}_n = [w f_{n-1}]/[f_{n-1}],
\end{eqnarray}
where $[F]$ of any function $F(x)$ is defined to be
\begin{eqnarray}\label{e1.29}
[F] = \int\limits_0^{\infty} \phi^2(x) F(x) dx.
\end{eqnarray}
Because of (\ref{e1.28}), at $x=0$
\begin{eqnarray}\label{e1.30}
f'_n(0) = 0.
\end{eqnarray}

In explicit form,
\begin{eqnarray}\label{e1.31}
f_1(x) &=& 1 -2\int\limits_x^{\infty} \phi^{-2}(y) dy \int\limits_y^{\infty} \phi^2(z)
(w(z) -  {\cal E}_1) dz, \\
{\cal E}_1 &=& \int\limits_0^{\infty} \phi^2(x)w(x)dx \bigg/
\int\limits_0^{\infty}\phi^2(x)dx,\\
f_2(x) &=& 1 -2\int\limits_x^{\infty} \phi^{-2}(y) dy \int\limits_y^{\infty} \phi^2(z)
(w(z) -  {\cal E}_2)f_1(z) dz, \\
{\cal E}_2 &=& \int\limits_0^{\infty} \phi^2(x)f_1(x) w(x) dx \bigg/
\int\limits_0^{\infty}\phi^2(x) f_1(x) dx,
\end{eqnarray}
etc. Because of (\ref{e1.28}), (\ref{e1.27}) can be written either as
\begin{eqnarray}\label{e1.35}
f_n(x) &=& 1 -2\int\limits_x^{\infty} \phi^{-2}(y) dy \int\limits_y^{\infty} \phi^2(z)
(w(z) -  {\cal E}_n)f_{n-1}(z) dz,
\end{eqnarray}
or in an equivalent form
\begin{eqnarray}\label{e1.36}
f_n(x) &=& f_n(0) -2\int\limits_0^x \phi^{-2}(y) dy \int\limits_0^y \phi^2(z)
(w(z) -  {\cal E}_n)f_{n-1}(z) dz
\end{eqnarray}
with
\begin{eqnarray}\label{e1.37}
{\cal E}_n &=& \int\limits_0^{\infty} \phi^2(x)f_{n-1}(x) w(x) dx \bigg/
\int\limits_0^{\infty}\phi^2(x) f_{n-1}(x) dx.
\end{eqnarray}

As we shall show, this iterative procedure is a convergent one.

\newpage

\section*{\bf 3. Some General Properties}
\setcounter{section}{3}
\setcounter{equation}{0}

For clarity, we list the following properties of $f_n$ and $ {\cal E}_n$ in the
form of several theorems.

\noindent
\underline{Theorem 3.1}~~~ At all $x>0$, the derivative of each $f_n(x)$ is negative;
i.e.,
\begin{eqnarray}\label{e2.1}
f_n'(x) \leq 0.
\end{eqnarray}
Thus,
\begin{eqnarray}\label{e2.2}
f_n(0) \geq f_n(x) \geq f_n(\infty)= 1.
\end{eqnarray}
\underline{Proof}

From (\ref{e1.35}), we have
\begin{eqnarray}\label{e2.3}
f_n'(x) = 2\phi^{-2}(x)\int\limits_x^{\infty} \phi^2(z)
(w(z) -  {\cal E}_n)f_{n-1}(z) dz
\end{eqnarray}
and , because of the equivalent form (\ref{e1.36}),
\begin{eqnarray}\label{e2.4}
f_n'(x) = -2\phi^{-2}(x)\int\limits_0^x \phi^2(z)
(w(z) -  {\cal E}_n)f_{n-1}(z) dz.
\end{eqnarray}
Assume that (\ref{e2.2}) holds for $f_n(x)$ when $n=m-1$. Let
\begin{eqnarray}\label{e2.5}
w(x)- {\cal E}_m = 0~~~~~~~~{\sf at}~~x=x_m.
\end{eqnarray}
Since $w'(x) < 0$ and because of (\ref{e1.37}), (\ref{e2.5}) has one and only one
solution. Furthermore, $w(x)-{\cal E}_m$ is negative for $x>x_m$ and positive for
$x<x_m$. It follows then $f_m'(x)<0$ for $x>x_m$ on account of (\ref{e2.3}), and
for $x<x_m$ because of (\ref{e2.4}). When $n=0$, $f_0(1)=1$ which satisfies
(\ref{e2.1})-(\ref{e2.2}). The theorem is proved by induction.

\noindent
\underline{Theorem 3.2}
\begin{eqnarray}\label{e2.6}
 {\cal E}_n >  {\cal E}_1
\end{eqnarray}
for all $n>1$.

\noindent
\underline{Proof}

From (\ref{e1.28}), we have
\begin{eqnarray}\label{e2.7}
{\cal E}_n - {\cal E}_1 &=& \frac{[w(x) f_{n-1}(x)]}{[f_{n-1}(x)]} -{\cal E}_1 \nonumber\\
                        &=& \frac{[(w(x)-{\cal E}_1) f_{n-1}(x)]}{[f_{n-1}(x)]}.
\end{eqnarray}
Let $x_1$ be the solution of
\begin{eqnarray}\label{e2.8}
w(x_1) - {\cal E}_1 = 0.
\end{eqnarray}
By using
\begin{eqnarray}\label{e2.9}
[w(x) - {\cal E}_1] = 0,
\end{eqnarray}
we may write (\ref{e2.7}) as
\begin{eqnarray}\label{e2.10}
{\cal E}_n - {\cal E}_1 = [(w(x)-{\cal E}_1) (f_{n-1}(x)-f_{n-1}(x_1))]/[f_{n-1}(x)].
\end{eqnarray}
Note that $w(x)-{\cal E}_1$ and $f_{n-1}(x)-f_{n-1}(x_1)$ are both negative for
$x>x_1$, and both positive for $x<x_1$. It follows then  ${\cal E}_n-{\cal E}_1$
is always positive.

\noindent
\underline{Theorem 3.3}
\begin{eqnarray}\label{e2.11}
1.~~~~~~~~~~~~~~~~~~~~~~~~~~~~~~~~{\cal E}_1 = \frac{1}{4} + \frac{9}{2^6}~\frac{1}{g} +
 \delta_1,~~~~~~~~~~~~~~~~~~~~~~~~~~~~~
\end{eqnarray}
where for $g$ sufficiently large
\begin{eqnarray}\label{e2.12}
0<\delta_1 = O(\frac{1}{g^2}) << 1.
\end{eqnarray}
~~~2.~~ Introduce
\begin{eqnarray}\label{e2.13}
I \equiv 2\int\limits_0^1 \phi^{-2}(y) dy \int\limits_0^y \phi^2(z)w(z) dz.
\end{eqnarray}
and
\begin{eqnarray}\label{e2.14}
J \equiv 2\int\limits_1^{\infty} \phi^{-2}(y) dy \int\limits_y^{\infty}
\phi^2(z)dz.
\end{eqnarray}
For $g$ sufficiently large,
\begin{eqnarray}\label{e2.15}
I \leq O(\frac{ln~\sqrt{g}}{g})<< 1,~~~~J \leq O(\frac{ln~\sqrt{g}}{g}) << 1.
\end{eqnarray}
The proofs of (\ref{e2.11}) - (\ref{e2.15}) are given in Appendix A, together with
the upper bounds of ${\cal E}_1$, $J$ and $I$.

It is useful to define
\begin{eqnarray}\label{e2.16}
\overline{f_n(0)},~~\overline{f_n(1)}~~{\sf and}~~\overline{{\cal E}_n}
\end{eqnarray}
to be respectively the maxima of $f_m(0)$, $f_m(1)$ and ${\cal E}_m$ for all
$m \leq n$.

\noindent
\underline{Theorem 3.4}
\begin{eqnarray}\label{e2.17}
\overline{f_n(0)} &<& \frac{\overline{f_n(1)}}{1-I},\\
\overline{f_n(1)} &<& \frac{1}{1-J \overline{{\cal E}_n}}~,\\
\overline{{\cal E}_n} &<& \overline{f_{n-1}(0)}~ {\cal E}_1 \leq
 \overline{f_n(0)}~ {\cal E}_1
\end{eqnarray}
and therefore
\begin{eqnarray}\label{e2.20}
\overline{{\cal E}_n} < \frac{{\cal E}_1}{1-I}~
\frac{1}{1-J \overline{{\cal E}_{n-1}}}
 \leq \frac{{\cal E}_1}{1-I}~\frac{1}{1-J \overline{{\cal E}_n}}  ~.
\end{eqnarray}
\underline{Proof}~~~From (\ref{e1.36}), we have
\begin{eqnarray}\label{e2.21}
f_m(0) &=& f_m(1) +2\int\limits_0^1 \phi^{-2}(y) dy \int\limits_0^y \phi^2(z)
(w(z) -  {\cal E}_m)f_{m-1}(z) dz\nonumber \\
       &<& f_m(1) +2\int\limits_0^1 \phi^{-2}(y) dy \int\limits_0^y \phi^2(z)
w(z) f_{m-1}(z) dz
\end{eqnarray}
On account of (\ref{e2.2}), $f_{m-1}(0) \geq f_{m-1}(z)$, (\ref{e2.21}) leads to
\begin{eqnarray}\label{e2.22}
f_m(0) &<& f_m(1) + f_{m-1}(0) ~2\int\limits_0^1 \phi^{-2}(y) dy
           \int\limits_0^y \phi^2(z)w(z) dz\nonumber \\
       &=& f_m(1) + f_{m-1}(0) I.
\end{eqnarray}
From the definition (\ref{e2.16}), $f_m(1) \leq \overline{f_n(1)}$,
$f_{m-1}(0) \leq \overline{f_{n-1}(0)}$ for all $m \leq n$. Thus (\ref{e2.22})
implies that for all $m \leq n$
\begin{eqnarray}
f_m(0) < \overline{f_n(1)} + \overline{f_{n-1}(0)} ~I,\nonumber
\end{eqnarray}
and therefore
\begin{eqnarray}
\overline{f_n(0)} < \overline{f_n(1)} + \overline{f_{n-1}(0)} ~I,\nonumber
\end{eqnarray}
which, on account of $\overline{f_{n-1}(0)} \leq \overline{f_n(0)}$, gives (\ref{e2.17}).

From (\ref{e2.6}) and (\ref{e2.11}), we have
\begin{eqnarray}\label{e2.23}
{\cal E}_n > {\cal E}_1 > \frac{1}{4}~.
\end{eqnarray}
For $x \geq 1$,  $w(x) \leq \frac{1}{4}$~,
\begin{eqnarray}\label{e2.24}
0 < {\cal E}_n - w(x) < {\cal E}_n
\end{eqnarray}
and, on account of (\ref{e1.35}) and (\ref{e2.2}),
\begin{eqnarray}
f_n(x) &<& 1 + 2{\cal E}_n \int\limits_x^{\infty} \phi^{-2}(y) dy
\int\limits_y^{\infty} \phi^2(z) f_{n-1}(z) dz,\nonumber\\
       &<& 1 + 2{\cal E}_n f_{n-1}(x) \int\limits_x^{\infty} \phi^{-2}(y) dy
\int\limits_y^{\infty} \phi^2(z) dz.\nonumber
\end{eqnarray}
Set $x=1$, we have
\begin{eqnarray}\label{e2.25}
f_n(1) < 1+ {\cal E}_n f_{n-1}(1) J~,
\end{eqnarray}
which leads to
\begin{eqnarray}
\overline{f_n(1)} < 1+ \overline{{\cal E}_n}~ \overline{f_{n-1}(1)}~ J
\leq 1+ \overline{{\cal E}_n}~ \overline{f_n(1)}~ J \nonumber
\end{eqnarray}
and (3.18).

Using (\ref{e1.28}) and (\ref{e2.2}), we obtain
\begin{eqnarray}
{\cal E}_n < \frac{f_{n-1}(0)[w]}{f_{n-1}(\infty)[1]} = f_{n-1}(0) ~{\cal E}_1,\nonumber
\end{eqnarray}
which gives (3.19). Combining (\ref{e2.17}) - (3.19), we derive (\ref{e2.20})
and complete the proof of Theorem 3.4.

Let $K$ be the lesser valued solution of the quadratic equation
\begin{eqnarray}\label{e2.26}
K = \frac{{\cal E}_1}{1-I} \frac{1}{1-JK}
\end{eqnarray}
i.e.,
\begin{eqnarray}\label{e2.27}
K = \frac{1}{2J}(1-\sqrt{1-4J{\cal E}_1/(1-I)}~).
\end{eqnarray}
For $g$ sufficiently large and using (\ref{e2.15}), we find
\begin{eqnarray}\label{e2.28}
K = \frac{{\cal E}_1}{1-I}(1+ \frac{J{\cal E}_1}{1-I}+\cdots);
\end{eqnarray}
therefore
\begin{eqnarray}\label{e2.29}
K>{\cal E}_1
\end{eqnarray}
and
\begin{eqnarray}\label{e2.30}
K \rightarrow {\cal E}_1~~~~{\sf when}~~~g \rightarrow \infty.
\end{eqnarray}
\underline{Theorem 3.5}
\begin{eqnarray}\label{e2.31}
1.& ~~~&\overline{{\cal E}_m} < K ~~~~{\sf for~any}~~~m \geq 1,\\
2.& ~~~&\lim_{m \rightarrow \infty} \overline{{\cal E}_m} \equiv
\overline{{\cal E}_{\infty}}~~~{\sf exists~and~is}~~ < K.
\end{eqnarray}
\underline{Proof}~~~Assume (\ref{e2.31}) holds for $m=n-1$, we have from (\ref{e2.20})
\begin{eqnarray}
\overline{{\cal E}_n} < \frac{{\cal E}_1}{1-I}~ \frac{1}{1-JK}=K.\nonumber
\end{eqnarray}
From (\ref{e2.29}) and the definition (\ref{e2.16}) of $\overline{{\cal E}_n}$, we find
\begin{eqnarray}
\overline{{\cal E}_1} \leq \overline{{\cal E}_2}\leq \cdots \leq
\overline{{\cal E}_n} < K \nonumber
\end{eqnarray}
for all $n$. Theorem 3.5 then follows.

\noindent
\underline{Theorem 3.6}~~~For $g$ sufficiently large,
\begin{eqnarray}
&1.&~~~ \lim_{n \rightarrow \infty} f_n(x) ~~~{\sf ~exists~for~all}~~ x \geq
0.\nonumber\\
&2.&~~~ \lim_{n \rightarrow \infty} {\cal E}_n ~~~{\sf ~exists}.\nonumber
\end{eqnarray}
The proof is given in Appendix B. With Theorem 3.6, we establish the convergence
of the new iterative solution.

As we shall see, from a practical point of view this new iterative process is also
quite effective in producing good approximate solutions of the quantum double-well
problem. The first iteration $f_1(x)$ will be examined in Appendix D.

\newpage

\section*{\bf 4. Generalization}
\setcounter{section}{4}
\setcounter{equation}{0}

In this and the following sections we will extend our analysis to the lowest odd
eigenfunction $\psi_{od}$. As an intermediate step we shall first construct a
solution $\psi_+$ of the Schroedinger equation
\begin{eqnarray}\label{e4.1}
(T+V-E_+)\psi_+ = 0
\end{eqnarray}
in the region
\begin{eqnarray}\label{e4.2}
x \geq 0,
\end{eqnarray}
where $H=T+V$ is the same Hamiltonian given by (\ref{e1.1}), and $\psi_+$ satisfies
the boundary conditions:
\begin{eqnarray}\label{e4.3}
{\sf at}~~~x=0,~~~~&&~~~\frac{\psi_+'}{\psi_+}(0) = \frac{\phi_+'}{\phi_+}(0) = g-1
\end{eqnarray}
and
\begin{eqnarray}\label{e4.4}
{\sf at}~~~x=\infty,~~~&&~~~\psi_+(\infty) =0.
\end{eqnarray}
As we shall see, the solution $\psi_+(x)$ can be derived from
\begin{eqnarray}\label{e4.5}
\phi_+(x) = e^{-g S_0(x)}\frac{2}{1+x}
\end{eqnarray}
given by (2.6), through the Green's function
\begin{eqnarray}\label{e4.6}
D_+ \equiv -2 \phi_+ \theta \phi_+^{-2} \theta \phi_+
\end{eqnarray}
and the integral equation
\begin{eqnarray}\label{e4.7}
\psi_+ = \phi_+ + D_+ (u-{\cal E}_+)\psi_+,
\end{eqnarray}
where $u=(1+x)^{-2}$ and $\theta$ are given by (\ref{e1.11}) and (\ref{e1.16}) and
${\cal E}_+$ is   related to $E_+$ of (\ref{e4.1}) by
\begin{eqnarray}\label{e4.8}
E_+ = g -{\cal E}_+,
\end{eqnarray}
analogous to (\ref{e1.21}). In this section we present the convergent iterative
process leading from $\phi_+$ to $\psi_+$. In the next section we will show how
$\psi_+$ can in turn lead to the lowest odd eigenstate $\psi_{od}$ of $H$.

As in (\ref{e1.23}) - (\ref{e1.37}), we write
\begin{eqnarray}\label{e4.9}
\psi_+ = \phi_+  f_+ ~,
\end{eqnarray}
where $f_+$ satisfies
\begin{eqnarray}\label{e4.10}
f_+ = 1 + \overline{D}_+ (u-{\cal E}_+)f_+
\end{eqnarray}
with
\begin{eqnarray}\label{e4.11}
\overline{D}_+ \equiv \phi_+^{-1} D_+ \phi_+ = -2 \theta \phi_+^{-2} \theta
\phi_+^2.
\end{eqnarray}
Differentiating (\ref{e4.9}), we have
\begin{eqnarray}\label{e4.12}
\frac{\psi_+'}{\psi_+} = \frac{\phi_+'}{\phi_+} + \frac{f_+'}{f_+}~.
\end{eqnarray}
The boundary condition (\ref{e4.3}) implies that at $x=0$, $f_+'(0)=0$, and
therefore
\begin{eqnarray}\label{e4.13}
\int\limits_0^{\infty} \phi_+^2 (u-{\cal E}_+)f_+ dx =0;
\end{eqnarray}
i.e.,
\begin{eqnarray}\label{e4.14}
{\cal E}_+ = \frac{\int\limits_0^{\infty} u \phi_+ \psi_+ dx}
{\int\limits_0^{\infty} \phi_+ \psi_+ dx}~.
\end{eqnarray}

Analogous to (\ref{e1.26}), we introduce the iterative sequences $\{f_{+,n}\}$ and
$\{{\cal E}_{+,n}\}$, with
\begin{eqnarray}
f_{+,0} = 1~~~~~~{\sf and}~~~~~{\cal E}_{+,0} = 0 \nonumber
\end{eqnarray}
when $n=0$. For $n \geq 1$,
\begin{eqnarray}\label{e4.15}
f_{+,n} = 1 + \overline{D}_+ (u-{\cal E}_{+,n})f_{+,n-1}
\end{eqnarray}
and
\begin{eqnarray}\label{e4.16}
{\cal E}_{+,n} = [u~f_{+,n-1}]_+~/~[f_{+,n-1}]_+ ~,
\end{eqnarray}
where for any function $F(x)$
\begin{eqnarray}\label{e4.17}
[F]_+ \equiv \int\limits_0^{\infty} \phi_+^2(x) F(x) dx.
\end{eqnarray}
Because of (\ref{e4.16}),
\begin{eqnarray}\label{e4.18}
f_{+,n}'(0) = 0.
\end{eqnarray}
Thus, (\ref{e4.15}) can be written either as
\begin{eqnarray}\label{e4.19}
f_{+,n}(x) = 1- 2\int\limits_x^{\infty} \phi_+^{-2}(y) dy
 \int\limits_y^{\infty} \phi_+^2(z) (u(z) - {\cal E}_{+,n}) f_{+,n-1}(z) dz
\end{eqnarray}
or equivalently,
\begin{eqnarray}\label{e4.20}
f_{+,n}(x) =f_{+,n}(0) - 2\int\limits_0^x \phi_+^{-2}(y) dy
 \int\limits_0^y \phi_+^2(z) (u(z) - {\cal E}_{+,n}) f_{+,n-1}(z) dz
\end{eqnarray}

Parallel to the theorems given in Section 3, we establish the corresponding
properties satisfied by $f_{+,n}$ and ${\cal E}_{+,n} $ in terms of Theorems 4.1-4.6
proven below:

\noindent
\underline{Theorem 4.1}~~~ At all $x>0$,
\begin{eqnarray}\label{e4.21}
f_{+,n}'(x) \leq 0
\end{eqnarray}
and therefore
\begin{eqnarray}\label{e4.22}
f_{+,n}(0) \geq f_{+,n}(x) \geq  f_{+,n}(\infty) =1.
\end{eqnarray}

\noindent
\underline{Theorem 4.2}
\begin{eqnarray}\label{e4.23}
{\cal E}_{+,n} \geq {\cal E}_{+,1}
\end{eqnarray}
for all $n>1$.

\noindent
\underline{Proof}~~~By following (\ref{e2.3}) - (\ref{e2.10}) and changing $f_n$,
${\cal E}_n$, $[F]$, $w$ to $f_{+,n}$, ${\cal E}_{+,n}$, $[F]_+$ and $u$
respectively, we can readily establish the above two theorems. Note that $u'<0$,
like $w'$.

\noindent
\underline{Theorem 4.3}
\begin{eqnarray}\label{e4.24}
1.~~~{\cal E}_{+,1} =\frac{1}{4} + \frac{9}{2^6}~\frac{1}{g} +a_1,
\end{eqnarray}
where
\begin{eqnarray}\label{e4.25}
a_1 < \frac{311}{2^6 g^2}
\end{eqnarray}
and, if we neglect $O(e^{-\frac{4}{3}g})$, $a_1$ is positive and $O(g^{-2})$.

\noindent
~~~~~~2. Introduce
\begin{eqnarray}\label{e4.26}
I_+ \equiv 2\int\limits_0^1 \phi_+^{-2}(y) dy
 \int\limits_0^y \phi_+^2(z) u(z) dz
\end{eqnarray}
and
\begin{eqnarray}\label{e4.27}
J_+ \equiv 2\int\limits_1^{\infty} \phi_+^{-2}(y) dy
 \int\limits_y^{\infty} \phi_+^2(z) dz.
\end{eqnarray}
We have
\begin{eqnarray}\label{e4.28}
I_+ < \frac{6}{g}(ln~\sqrt{\frac{\pi g}{3}}~+1 )
\end{eqnarray}
and
\begin{eqnarray}\label{e4.29}
J_+ < \frac{1}{2g}ln~(e+2e\sqrt{2\pi g}~).
\end{eqnarray}

\noindent
\underline{Proof}~~~To prove this theorem, we need several formulas, already
established in Appendix A for the proof of Theorem 3.3 in Section 3.
Eq. (\ref{e4.24}) is identical to (A.2), and (A.3) gives (\ref{e4.25}).

From (\ref{e4.26}), we see that
\begin{eqnarray}\label{e4.30}
I_+ = 2\int\limits_0^1 dy
 \int\limits_0^y dz~e^{-2g(S_0(z)-S_0(y))} \frac{(1+y)^2}{(1+z)^4}~,
\end{eqnarray}
which is $\frac{1}{4}$ of the righthand side of (A.60). By following (A.61) -
(A.64), we see that $I_+$ is less than $\frac{1}{4}$ of the bound (A.65), which
leads to (\ref{e4.28}).

According to (\ref{e1.8}), $\phi(x) \propto \phi_+(x)$ for $x>1$. Thus $J_+$ above
is the same  $J$ given by (\ref{e2.14}). Its upper bound (\ref{e4.29}) is the same
one given by (A.39.2). Theorem 4.3 is then proved.

As in (\ref{e2.16}), define
\begin{eqnarray}\label{e4.31}
\overline{f_{+,n}(0)},~~\overline{f_{+,n}(1)}~~{\sf and}~~\overline{{\cal E}_{+,n}}
\end{eqnarray}
to be respectively the maxima of $f_{+,m}(0),~~f_{+,m}(1)$ and ${\cal E}_{+,m}$ for
all $m \leq n$.

\noindent
\underline{Theorem 4.4}
\begin{eqnarray}\label{e4.32}
\overline{f_{+,n}(0)} &<& \frac{\overline{f_{+,n}(1)}}{1-I_+}~,\\
\overline{f_{+,n}(1)} &<& \frac{1}{1-J_+\overline{{\cal E}_{+,n}}}~,\\
\overline{{\cal E}_{+,n}} &<& \overline{f_{+,n-1}(0)} {\cal E}_{+,1}
\leq \overline{f_{+,n}(0)} {\cal E}_{+,1}
\end{eqnarray}
and consequently
\begin{eqnarray}\label{e4.35}
\overline{{\cal E}_{+,n}} < \frac{{\cal E}_{+,1}}{1-I_+}~
\frac{1}{1-J_+ \overline{{\cal E}_{+,n-1}}}
\leq \frac{{\cal E}_{+,1}}{1-I_+}~
\frac{1}{1-J_+ \overline{{\cal E}_{+,n}}}~.
\end{eqnarray}
Analogous to (\ref{e2.26}) - (\ref{e2.27}), introduce
\begin{eqnarray}\label{e4.36}
K_+ \equiv \frac{1}{2J_+}(1-\sqrt{1-4J_+{\cal E}_{+,1}/(1-I_+)}~),
\end{eqnarray}
which is also the lesser valued solution of the quadratic equation
\begin{eqnarray}\label{e4.37}
K_+ = \frac{{\cal E}_{+,1}}{1-I_+}~\frac{1}{1-J_+ K_+}~.
\end{eqnarray}
Thus, $K_+ > {\cal E}_{+,1}$ and $\rightarrow {\cal E}_{+,1}$ when $g \rightarrow
\infty$.

\noindent
\underline{Theorem 4.5}
\begin{eqnarray}\label{e4.38}
&1.&~~~ \overline{{\cal E}_{+,m}} < K_+ ~~~{\sf ~for~all}~~ m \geq 1 \\
&2.&~~~ \lim_{m \rightarrow \infty} \overline{{\cal E}_{+,m}}
 ~~~{\sf ~exists~and~is}~~~ < K_+~.
\end{eqnarray}

\noindent
\underline{Theorem 4.6}~~~For $g$ sufficiently large,
\begin{eqnarray}\label{e4.40}
&1.&~~~ \lim_{n \rightarrow \infty} f_{+,n}(x) \equiv f_+(x)
 ~~~{\sf ~exists~for~all}~~ x \geq 0.\\
&2.&~~~ \lim_{n \rightarrow \infty} {\cal E}_{+,n} \equiv {\cal E}_+ ~~~{\sf ~exists}.
\end{eqnarray}
The proofs of these theorems are identical to those of Theorems 3.4, 3.5 and 3.6,
provided we replace $f_n,~{\cal E}_n,~w,~I,~J$ and $K$ by $f_{+,n},~{\cal E}_{+,n},
~u,~I_+,~J_+$ and $K_+$. Thus, we establish the convergence of the new iterative
solution for $\psi_+$.

It is straightforward to verify that, similar to (\ref{e4.21}) and (\ref{e4.22}),
for $x>0$
\begin{eqnarray}\label{e4.42}
f_+'(x) < 0
\end{eqnarray}
and therefore,
\begin{eqnarray}\label{e4.43}
f_+(0) > f_+(x) > f_+(\infty) =1.
\end{eqnarray}
Furthermore,
\begin{eqnarray}\label{e4.44}
f_+(0) < (1-L)^{-1},
\end{eqnarray}
where $L < O(\frac{1}{g}~ln~\sqrt{g})$ is given by
\begin{eqnarray}\label{e4.45}
L = 1-(1 - I_+) (1 - J_+ {\cal E}_+) = I_+ + J_+ {\cal E}_+ - I_+ J_+ {\cal E}_+
\end{eqnarray}
with bounds of $I_+,~J_+$ and ${\cal E}_+$ given by (\ref{e4.28}), (\ref{e4.29}) and
${\cal E}_+ < K_+$, on account of (\ref{e4.38}). In addition, as mentioned before,
after (\ref{e4.12}), at $x=0$
\begin{eqnarray}\label{e4.46}
f_+'(0) = 0.
\end{eqnarray}

  Eq.(4.42) gives an upper bound for $f_+'(x)$. It is also useful to
set a lower bound for $f_+'(x)$. Differentiating (4.10), we have
\begin{eqnarray}\label{e4.47}
f_+'(x)=2\phi_+^{-2}(x) \int\limits_x^{\infty}\phi_+^2(z)
(u(z)-{\cal E}_+)f_+(z)dz.
\end{eqnarray}
Because of (4.13), the same expression can also be written as
\begin{eqnarray}\label{e4.48}
f_+'(x)=-2\phi_+^{-2}(x) \int\limits_0^x \phi_+^2(z)
(u(z)-{\cal E}_+)f_+(z)dz.
\end{eqnarray}
Since $u(z)=(1+z)^{-2}>0$, (4.47) gives
\begin{eqnarray}\label{e4.49}
f_+'(x) &>& -2{\cal E}_+ \phi_+^{-2}(x) \int\limits_x^{\infty}\phi_+^2(z)f_+(z)dz \nonumber\\
        &>& -2{\cal E}_+ f_+(x)  \phi_+^{-2}(x) \int\limits_x^{\infty}\phi_+^2(z)dz
\end{eqnarray}
From (4.23)-(4.25),(A.4)-(A.5) in Appendix A and neglecting
$O(e^{-\frac{4}{3}g})$ we have ${\cal E}_+>\frac{1}{4}~$. The
replacement of ${\cal E}_+$ by $\frac{1}{4}$ in (4.48) yields the
inequality
\begin{eqnarray}\label{e4.50}
f_+'(x) &>& -2 \phi_+^{-2}(x) \int\limits_0^x \phi_+^2(z)(u(z)-\frac{1}{4})f_+(z)dz\nonumber\\
        &>& -2 f_+(0)\phi_+^{-2}(x) \int\limits_0^x \phi_+^2(z)(u(z)-\frac{1}{4})
        dz .
\end{eqnarray}

  For $x>1$, $\phi_+^2(x)=e^{-2gS_0(x)}(\frac{2}{1+x})^2 < e^{-2gS_0(x)}$.
Also, according to (A.33)-(A.36) proven in Appendix A,
\begin{eqnarray}\label{e4.51}
j(x) \equiv e^{2gS_0(x)}
\int\limits_x^{\infty} e^{-2gS_0(z)}dz < \frac{C}{(1+x)^2}
\end{eqnarray}
where
\begin{eqnarray}\label{e4.52}
C=\sqrt{\frac{2\pi}{g}}\Big( 1 + \frac{3}{2\sqrt{2\pi g}} + \cdots \Big)~.
\end{eqnarray}
Consequently, for $x>1$, (4.49) leads to
\begin{eqnarray}\label{e4.53}
f_+'(x)> -\frac{1}{2}{\cal E}_+ f_+(0)C~.
\end{eqnarray}

  For $0<z<x<1$, we have
\begin{eqnarray}
S_0(z)-S_0(x) &=& (x-z)(1-\frac{1}{3}(x^2+xz+z^2)) \nonumber\\
              &>& (x-z)(x-\frac{1}{3}(x+z+z))     \nonumber
\end{eqnarray}
and
\begin{eqnarray}
(\frac{1+x}{1+z})^2 (\frac{1}{(1+z)^2} - \frac{1}{4}) <
\frac{3}{4}(1+x)^2 < 3 \nonumber
\end{eqnarray}
which yield
\begin{eqnarray}\label{e4.54}
\phi_+^{-2}(x) \phi_+^2(z)(u(z)-\frac{1}{4}) =
e^{2g(S_0(x)-S_0(z))}(\frac{1+x}{1+z})^2 (\frac{1}{(1+z)^2} -
\frac{1}{4}) < 3e^{-\frac{4}{3}g(x-z)^2}~.
\end{eqnarray}
Thus, from (4.50) and for $x<1$, we derive
\begin{eqnarray}\label{e4.55}
f_+'(x) > -6f_+(0) \int\limits_0^x e^{-\frac{4}{3}g(x-z)^2} dz
> -\frac{3}{2} \sqrt{\frac{3\pi}{g}} f_+(0)~.
\end{eqnarray}

  Combining (4.52)-(4.53) with (4.55) and for $g$ sufficiently
large we have, at all $x$,
\begin{eqnarray}\label{e4.56}
f_+'(x) > -\frac{3}{2}\sqrt{\frac{3\pi}{g}} f_+(0),
\end{eqnarray}
which is a convenient (though not the best) lower bound.

In Appendix D, we will give an analysis of the first iterative solution
$f_{+,1}(x)$.


\newpage
\section*{\bf 5. The Lowest Odd Wave Function}
\setcounter{section}{5}
\setcounter{equation}{0}

\noindent
5.1.~~~Formulation

The lowest odd wave function $\psi_{od}$ of the double-well potential
$V=\frac{g^2}{2}(x^2-1)^2$ satisfies a similar Schroedinger equation as (\ref{e1.1}):
\begin{eqnarray}\label{e5.1}
(T+V-E_{od}) \psi_{od} = 0.
\end{eqnarray}
with
\begin{eqnarray}\label{e5.2}
\psi_{od}(-x) =- \psi_{od}(x),
\end{eqnarray}
and therefore, $\psi_{od}(0)=0$; besides at $x=0$, $\psi_{od}$ has no other node. In
this section, we give the derivation of $\psi_{od}(x)$ through an alternative
convergent iterative
procedure which differs from those of the previous sections. In this new approach,
as we shall see, each successive iteration yields an improvement by a factor
\begin{eqnarray}\label{e5.3}
\epsilon \equiv e^{-\frac{4}{3}g}
\end{eqnarray}
that is much smaller than the previous $O(g^{-1})$ factor.

We begin with the function $\psi_+(x)$ of (\ref{e4.1}), that has been derived in the
last section. Recall that while $\psi_+(x)$ satisfies the differential equation
$(T+V-E_+) \psi_+ = 0$, it is not an eigenstate of the Hamiltonian $H=T+V$, since at
$x=-\infty$, $\psi_+$ diverges. However, as will be shown below, it is possible to
construct a good trial function $\chi (x)$, odd in $x$ and well behaved everywhere,
including $x= \pm \infty$. For $x\ge 0$, we write
\begin{eqnarray}\label{e5.4}
\chi(x) \equiv \psi_+(x)-\psi_-(x),
\end{eqnarray}
where
\begin{eqnarray}\label{e5.5}
\psi_-(x) \equiv 2 \gamma \psi_+(x) \int\limits^{\infty}_x \psi_+^{-2}(y)dy
\int\limits^{\infty}_y \psi_+^2(z) dz
\end{eqnarray}
with the constant $\gamma$ chosen to make  $\chi$ vanish at $x=0$,
\begin{eqnarray}\label{e5.6}
\chi(0) = \psi_+(0)-\psi_-(0)=0.
\end{eqnarray}
For $x<0$, we set
\begin{eqnarray}\label{e5.7}
\chi(x) = - \chi(-x)
\end{eqnarray}
Because of (\ref{e5.6}), $\gamma$ is a positive constant given by
\begin{eqnarray}\label{e5.8}
\gamma^{-1} = 2 \int\limits^{\infty}_0 \psi_+^{-2}(y)dy
\int\limits^{\infty}_y \psi_+^2(z) dz.
\end{eqnarray}

It can be readily verified that at $x=0$, both $\chi(x)$ and $\chi'(x)$ are
continuous, but not $\chi''(x)$. Furthermore, by differentiating (\ref{e5.5}), we
have
\begin{eqnarray}\label{e5.9}
(T+V-E_+)\psi_- = - \gamma \psi_+.
\end{eqnarray}
Define $\nu(x)$ by
\begin{eqnarray}\label{e5.10}
\nu(x) \chi(x) = - \gamma \psi_+(x)
\end{eqnarray}
for $x>0$, and $\nu(x) = \nu(-x)$ for $x<0$.

\noindent
\underline{Theorem 5.1}

(i) $\chi$ is the lowest odd eigenstate of the Hamiltonian $T+V+\nu$, with
eigenvalue $E_+$; i.e.,
\begin{eqnarray}\label{e5.11}
(T+V+\nu-E_+)\chi = 0.
\end{eqnarray}

(ii)
\begin{eqnarray}\label{e5.12}
\nu(x) < 0,
\end{eqnarray}
and for $x>0$
\begin{eqnarray}\label{e5.13}
\nu'(x) > 0.
\end{eqnarray}
(Since $\nu(x)$ is even in $x$, $\nu'(x) < 0$ for $x$ negative.)

\noindent
\underline{Proof}

(i) Write (\ref{e5.4}) as
\begin{eqnarray}\label{e5.14}
\chi = \psi_+(1-\frac{\psi_-}{\psi_+}).
\end{eqnarray}
From (\ref{e5.5}),
\begin{eqnarray}\label{e5.15}
(\frac{\psi_-}{\psi_+})'<0.
\end{eqnarray}
Since at $x=0$, $\psi_- = \psi_+$, we have for $x>0$
\begin{eqnarray}\label{e5.16}
\psi_-(x)<\psi_+(x),
\end{eqnarray}
and therefore
\begin{eqnarray}\label{e5.17}
\chi(x) \ge 0,
\end{eqnarray}
in which $\chi(x)=0$ occurs only at $x=0$ (and $x \rightarrow \infty$). In
(\ref{e5.15})-(\ref{e5.17}), $x$ is restricted to $x\ge 0$. (In the following,
unless specified, the same restriction applies.)

From (\ref{e5.7}) and (\ref{e5.9})-(\ref{e5.10}), we see that $\chi(x)$
satisfies the Schroedinger
equation (\ref{e5.11}) at all $x$. Since $\chi(x)$ has only one node at $x=0$, it
is the lowest odd eigenstate of the Hamiltonian $T+V+\nu$.

(ii) From (\ref{e5.10}), we have $\nu<0$.

Multiply (\ref{e4.1}) and (\ref{e5.11}) on the left by $\chi$ and $\psi_+$
respectively, and take their difference. We obtain
\begin{eqnarray}
(\chi'\psi_+-\psi_+'\chi)' = 2\nu \chi \psi_+ = -2 \gamma \psi_+^2.\nonumber
\end{eqnarray}
Consequently
\begin{eqnarray}\label{e5.18}
\big(\frac{\psi_+(x)}{\chi(x)}\big)'= -2 \gamma \chi^{-2}(x)
 \int\limits^{\infty}_x \psi_+^2(y)dy.
\end{eqnarray}
Since the left side is $-\nu'/\gamma$, we derive (\ref{e5.13}) and complete the
proof for Theorem 5.1.

\noindent
\underline{Theorem 5.2}~~~The constant $\gamma$ satisfies
\begin{eqnarray}\label{e5.19}
\gamma < 4g \sqrt{\frac{2g}{\pi}}~e^{-\frac{4}{3}g}(1+\alpha_{\gamma})
\end{eqnarray}
and
\begin{eqnarray}\label{e5.20}
\gamma > 4g \sqrt{\frac{2g}{\pi}}~e^{-\frac{4}{3}g}(1-\beta_{\gamma}),
\end{eqnarray}
where
\begin{eqnarray}\label{e5.21}
\alpha_{\gamma}=O(\frac{1}{g}\ln~\sqrt{g})~~~{\sf and}~~~
\beta_{\gamma}=O(\frac{1}{g^{1/3}}),
\end{eqnarray}
so that for $g$ sufficiently large
\begin{eqnarray}\label{e5.22}
\gamma \cong 4g \sqrt{\frac{2g}{\pi}}~e^{-\frac{4}{3}g}.
\end{eqnarray}

The explicit form and the derivation of these bounds $\alpha_{\gamma}$ and
$\beta_{\gamma}$ will be given in Appendix C. Here we only discuss how to obtain the
approximate formula (\ref{e5.22}).

In (\ref{e5.8}), because of (\ref{e4.43})-(\ref{e4.44}), for $g$ sufficiently large
\begin{eqnarray}
\psi_+(x) = f_+(x) \phi_+(x) \cong \phi_+(x).\nonumber
\end{eqnarray}
Hence, (\ref{e5.8}) can be approximated by
\begin{eqnarray}\label{e5.23}
\gamma^{-1} \cong 2 \int\limits^{\infty}_0 \phi_+^{-2}(y)dy
\int\limits^{\infty}_y \phi_+^2(z) dz.
\end{eqnarray}
According to (2.6), the $z$-integration is dominated by the region
$z=1+O(\frac{1}{\sqrt{g}})$, where
\begin{eqnarray}\label{e5.24}
\phi_+^2(z) \cong e^{-2g(z-1)^2}
\end{eqnarray}
is near its peak, whereas the $y$-integration is largely determined by the region
$y$ near $0$, and $O(\frac{1}{g})$, where
\begin{eqnarray}\label{e5.25}
\phi_+^{-2}(y) =(\frac{1+y}{2})^2 e^{\frac{2}{3}gy^3-2gy+\frac{4}{3}g}
\cong \frac{1}{4}e^{-2gy+\frac{4}{3}g}.
\end{eqnarray}
Thus, we may futher approximate (\ref{e5.23}) by
\begin{eqnarray}
\gamma^{-1} \cong 2 \int\limits^{\infty}_0 \frac{1}{4}e^{-2gy+\frac{4}{3}g}
 \int\limits^{\infty}_ {- \infty} e^{-2g(z-1)^2}
= \frac{1}{4g}e^{\frac{4}{3}g}\sqrt{\frac{\pi}{2g}}\nonumber
\end{eqnarray}
which gives (\ref{e5.22}). Therefore, on account of (\ref{e5.3}),
$\gamma = O(\epsilon)$, and so is $\nu(x)$.

Because $\chi$ and $\psi_{od}$ are both the lowest odd eigenstate of two closely
related Hamiltonians which differ only by
\begin{eqnarray}\label{e5.26}
\nu(x) = O(\epsilon),
\end{eqnarray}
the iterative approach for the derivation of $\psi_{od}$ by using $\chi$ as the
zero$^{th}$ approximation should be an effective one, as will be shown. (In the
derivation, $\nu(x)$ appears only as part of the product $\nu \chi$; therefore its
effect is always small, $O(\epsilon)$, even though $\nu(x)$ by itself is singular
at $x=0$.)

The Green's function associated with $\chi$ is
\begin{eqnarray}\label{e5.27}
G \equiv -2 \chi \theta \chi^{-2} \theta \chi,
\end{eqnarray}
where, as before, $\theta$ is given by (\ref{e1.16}). Because of (\ref{e5.11}), $G$
satisfies
\begin{eqnarray}\label{e5.28}
(T+V+\nu -E_+)G = 1.
\end{eqnarray}
Therefore, the eigenstate $\psi_{od}$, defined by (\ref{e5.1}), is the solution of
\begin{eqnarray}\label{e5.29}
\psi_{od} = \chi + G(\nu +\Delta_{od})\psi_{od},
\end{eqnarray}
where
\begin{eqnarray}\label{e5.30}
\Delta_{od} = E_{od} - E_+;
\end{eqnarray}
i.e.
\begin{eqnarray}\label{e5.31}
\psi_{od}(x) =\chi(x) - 2 \chi(x) \int\limits^{\infty}_x \chi^{-2}(y)dy
\int\limits^{\infty}_y (\nu(z) + \Delta_{od})\chi(z) \psi_{od}(z)dz.
\end{eqnarray}

At $x=0$, in order that the above equation agrees with $ \psi_{od}(0)=\chi(0)=0$,
the integrand must satisfy
\begin{eqnarray}
\lim_{y \rightarrow 0} \chi^{-2}(y)
\int\limits^{\infty}_y (\nu(z) + \Delta_{od})\chi(z) \psi_{od}(z)dz = ~{\sf finite},
\nonumber
\end{eqnarray}
which implies
\begin{eqnarray}\label{e5.32}
\int\limits^{\infty}_0 (\nu(z) + \Delta_{od})\chi(z) \psi_{od}(z)dz = 0
\end{eqnarray}
and, on account of (\ref{e5.10}),
\begin{eqnarray}\label{e5.33}
\Delta_{od} = \gamma \frac{\int\limits^{\infty}_0 \psi_+(x) \psi_{od}(x)dx}
{\int\limits^{\infty}_0 \chi(x) \psi_{od}(x)dx}~>0.
\end{eqnarray}

The positivity of $\Delta_{od} = E_{od} -E_+$ is to be expected. For $x\ge 0$, let
$\psi$ be the solution of the second order differential equation
\begin{eqnarray}\label{e5.34}
(T+V-E)\psi = 0.
\end{eqnarray}
Assume that $\psi$ has no node when $x>0$, and it satisfies the boundary conditions:
\begin{eqnarray}\label{e5.35}
&&\frac{\psi'}{\psi} = p \ge 0~~~{\sf at}~~x=0\nonumber\\
{\sf and} ~~~~~~~~~~~~~~~~~~&&\\
&&\psi=0~~~~~~~~~~~~~~{\sf at}~~x=\infty.\nonumber
\end{eqnarray}
When the slope $p=0$, $\psi=\psi_{ev}$ and $E=E_{ev}$. As $p$ increases, so does
$E$. When $p=\phi_+'(0)/\phi_+(0)$, $\psi=\psi_+$ and $E=E_+>E_{ev}$, and when
$p=\infty$,  $\psi=\psi_{od}$ and $E=E_{od}$, with
\begin{eqnarray}\label{e5.36}
E_{od}=E_+ + \Delta_{od} > E_+.
\end{eqnarray}

Write
\begin{eqnarray}\label{e5.37}
\psi_{od}(x) = \chi(x)k(x).
\end{eqnarray}
On account of (\ref{e5.31}), $k(x)$ satisfies
\begin{eqnarray}\label{e5.38}
k(x) =1 - 2 \int\limits^{\infty}_x \chi^{-2}(y)dy
\int\limits^{\infty}_y (\nu(z) + \Delta_{od})\chi^2(z)k(z)dz.
\end{eqnarray}
Define
\begin{eqnarray}\label{e5.39}
\overline{G} = -2 \theta \chi^{-2} \theta \chi^2.
\end{eqnarray}
Eq.(\ref{e5.38}) can also be written as
\begin{eqnarray}\label{e5.40}
k =1 +\overline{G}(\nu + \Delta_{od})k.
\end{eqnarray}
As in (\ref{e1.26})-(\ref{e1.28}) and (\ref{e4.15})-(\ref{e4.17}), $k(x)$, and
therefore $\psi_{od}(x)$, will be solved by an iterative procedure.

Introduce the sequences $\{k_n\}$ and $\{\Delta_n\}$, with
\begin{eqnarray}\label{e5.41}
k_0 =1~~~~{\sf and}~~~~\Delta_0=0.
\end{eqnarray}
For $n\ge 1$,
\begin{eqnarray}\label{e5.42}
k_n =1 +\overline{G}(\nu + \Delta_n)k_{n-1}
\end{eqnarray}
and
\begin{eqnarray}\label{e5.43}
\Delta_n =- \{\nu k_{n-1}\} \big/ \{ k_{n-1}\},
\end{eqnarray}
where for any function $F(x)$
\begin{eqnarray}\label{e5.44}
\{ F\} \equiv \int\limits^{\infty}_0 \chi^2(x) F(x) dx.
\end{eqnarray}
Thus, (\ref{e5.42}) can be written as either
\begin{eqnarray}\label{e5.45}
k_n(x) =1 - 2 \int\limits^{\infty}_x \chi^{-2}(y)dy
\int\limits^{\infty}_y (\nu(z) + \Delta_n)\chi^2(z)k_{n-1}(z)dz,
\end{eqnarray}
or
\begin{eqnarray}\label{e5.46}
k_n(x) =k_n(0) - 2 \int\limits^x_0 \chi^{-2}(y)dy
\int\limits^y_0 (\nu(z) + \Delta_n)\chi^2(z)k_{n-1}(z)dz.
\end{eqnarray}

\noindent
5.2 ~~Convergence

\noindent
\underline{Theorem 5.3}

(i) For $n \ge 0$ and $x>0$, we have
\begin{eqnarray}\label{e5.47}
k'_n(x) \ge 0,
\end{eqnarray}
and therefore
\begin{eqnarray}\label{e5.48}
k_n(0) \leq k_n(x)\leq k_n(\infty)= 1.
\end{eqnarray}
Furthermore,
\begin{eqnarray}\label{e5.49}
k_n(0) \geq {\cal K} >0
\end{eqnarray}
where ${\cal K} $ is independent of $n$.

(ii) For $n>1$,
\begin{eqnarray}\label{e5.50}
\Delta_1>\Delta_n>0.
\end{eqnarray}

\noindent
\underline{Proof}~~~ Assume that (\ref{e5.47})-(\ref{e5.50}) hold for $n=m-1$.
From (\ref{e5.45}) and (\ref{e5.46}), we see that for $n=m$,
\begin{eqnarray}\label{e5.51}
k'_m(x)&=& 2\chi^{-2}(x)\int\limits^{\infty}_x(\nu(z)+\Delta_m)\chi^2(z)k_{m-1}(z)dz
\nonumber\\
       &=& -2\chi^{-2}(x)\int\limits^x_0(\nu(z)+\Delta_m)\chi^2(z)k_{m-1}(z)dz.
\end{eqnarray}
By assumption, $k_{m-1}(x)>0$. From (\ref{e5.12}), $\nu<0$; therefore, (\ref{e5.43})
gives $\Delta_m>0$.

Let
\begin{eqnarray}\label{e5.52}
\nu(x)+\Delta_m=0~~~{\sf at}~~x=x_m.
\end{eqnarray}
Since, according to (\ref{e5.13}), $\nu'>0$, we see that (\ref{e5.52}) has only one
solution. Furthermore, $\nu(x) + \Delta_m$ is positive for $x>x_m$, and negative for
$x<x_m$. Thus, by using either the upper, or the lower equation in
(\ref{e5.51}), we find $k'_m(x)>0$ when $x>0$.

From (\ref{e5.43}),
\begin{eqnarray}\label{e5.53}
-\Delta_m + \Delta_1&=&\frac{\{\nu(x)k_{m-1}(x)\}}{\{k_{m-1}(x)\}} + \Delta_1\nonumber\\
&=&\frac{\{(\nu(x)+\Delta_1)k_{m-1}(x)\}}{\{k_{m-1}(x)\}}~.
\end{eqnarray}
According to (\ref{e5.52}), $\nu(x)+\Delta_1=0$ at $x=x_1$. Since $k_0=1$, we have
$\{\nu(x)+\Delta_1\}=0$. Write (\ref{e5.53}) as
\begin{eqnarray}\label{e5.54}
-\Delta_m + \Delta_1=\frac{\{(\nu(x)+\Delta_1)(k_{m-1}(x)-k_{m-1}(x_1))\}}{\{k_{m-1}(x)\}}~.
\end{eqnarray}
When $x>x_1$, both $\nu(x)+\Delta_1$ and $k_{m-1}(x)-k_{m-1}(x_1)$ are positive; when
$x<x_1$, both negative. Therefore, for $m>1$, $\Delta_m<\Delta_1$.
To complete the induction, we still have to prove that when $n=m$, (\ref{e5.49}) is true,
also (\ref{e5.47})-(\ref{e5.49}) hold for $n=0$ and (\ref{e5.50}) is valid for $n=2$.

By assumption, $0<k_{m-1}(x)<1$. Since $\nu(x)$ is negative, and for $m>1$, $\Delta_m$ is
positive but $<\Delta_1$, we have
\begin{eqnarray}\label{e5.55}
\int\limits^{\infty}_y(\nu(z)+\Delta_m)\chi^2(z)k_{m-1}(z)dz &<&
\int\limits^{\infty}_y \Delta_m\chi^2(z)dz\nonumber\\
& <& \Delta_1 \int\limits^{\infty}_y \chi^2(z)dz;
\end{eqnarray}
therefore at $x=1$, (\ref{e5.45}) implies
\begin{eqnarray}\label{e5.56}
k_m(1) > 1-2\Delta_1 \int\limits^{\infty}_1 \chi^{-2}(y)dy
\int\limits^{\infty}_y \chi^2(z)dz.
\end{eqnarray}
From (\ref{e5.46}), we have
\begin{eqnarray}
k_m(0) &=& k_m(1) + 2\int\limits^1_0 \chi^{-2}(y)dy \int\limits^y_0(\nu(z) +
\Delta_m)\chi^2(z)k_{m-1}(z)dz\nonumber\\
&>& k_m(1) + 2\int\limits^1_0 \chi^{-2}(y)dy \int\limits^y_0\nu(z)\chi^2(z)dz,\nonumber
\end{eqnarray}
which, on account of $\nu\chi=-\gamma\psi_{+}$, gives
\begin{eqnarray}\label{e5.57}
k_m(0) > k_m(1) - 2 \gamma\int\limits^1_0 \chi^{-2}(y)dy
\int\limits^y_0\chi(z)\psi_{+}(z)dz.
\end{eqnarray}

Define
\begin{eqnarray}\label{e5.58}
{\cal I} \equiv 2\int\limits^1_0 \chi^{-2}(y)dy \int\limits^y_0\chi(z)\psi_{+}(z)dz
\end{eqnarray}
and
\begin{eqnarray}\label{e5.59}
{\cal J}^+ \equiv 2\int\limits^{\infty}_1 \chi^{-2}(y)dy \int\limits^{\infty}_y\chi^2(z)dz.
\end{eqnarray}
We see that from (\ref{e5.56})-(\ref{e5.57}),
\begin{eqnarray}\label{e5.60}
k_m(0) > 1 - \gamma \Big[ {\cal I} + {\cal J}^+ \frac{\Delta_1}{\gamma} \Big]  \equiv {\cal K}.
\end{eqnarray}

In Appendix C, an upper bound will be given for
${\cal I} + {\cal J}^+\frac{\Delta_1}{\gamma}$,
which is $O(\epsilon^0)$. (See (C.86), (C.87), (C.107) and (C.118).)
Since $\gamma$ satisfies
(\ref{e5.19})-(\ref{e5.22}), we have ${\cal K}=1-O(\epsilon) > 0$.
When $n=0$, $k_0=1$ which satisfies (\ref{e5.47})-(\ref{e5.49}). This leads to
(\ref{e5.47})-(\ref{e5.49}) being valid also
for $n=1$; in addition, $\Delta_2 > \Delta_1$ (since for $m=2$, the proof
(\ref{e5.53})-(\ref{e5.54}) depends only on $k_1$ satisfying
(\ref{e5.47})-(\ref{e5.49})). Theorem 5.3 is then established.

\noindent
\underline{Theorem 5.4}
\begin{eqnarray}\label{e5.61}
1.~~&& \lim_{n\rightarrow \infty} k_n(x) = k(x)~~~{\sf exists~for~all~}~~x \\
2.~~&& \lim_{n\rightarrow \infty} \Delta_n = \Delta_{od}~~~~~{\sf exists}
\end{eqnarray}
\underline{Proof}~~~~The proof is parallel to that of Theorem 3.6, or Theorem 4.6.

Introduce
\begin{eqnarray}\label{e5.63}
\kappa^n_j(x) \equiv k_{n+j}(x) -k_n(x)
\end{eqnarray}
and
\begin{eqnarray}\label{e5.64}
\overline{\kappa}~^n_j \equiv \max |\kappa_j^n(x)|~~{\sf over~all~}~~x.
\end{eqnarray}
As we shall see,

\noindent
\underline{Lemma}
\begin{eqnarray}\label{e5.65}
\overline{\kappa}~^n_j/\overline{\kappa}~^{n-1}_j \le r,
\end{eqnarray}
where $r=O(\epsilon)$ is finite and independent of $n$ and $j$.

In the following, from (\ref{e5.66})-(\ref{e5.91}) we first give the proof of the
lemma.

According to (\ref{e5.42}) and (\ref{e5.63}), we have
\begin{eqnarray}\label{e5.66}
\kappa^n_j = \overline{G}\Big((\nu + \Delta_{n+j}) k_{n+j-1} -(\nu + \Delta_n)
k_{n-1}\Big) \equiv  \overline{G}(Q_1 + Q_2 + Q_3),
\end{eqnarray}
where
\begin{eqnarray}\label{e5.67}
Q_1 &=& ( \Delta_{n+j} - \Delta_n) k_{n+j-1}\\
Q_2 &=&  \Delta_n( k_{n+j-1} - k_{n-1}) \\
Q_3 &=& \nu (k_{n+j-1} - k_{n-1}).
\end{eqnarray}

  Define for $x \ge 1$,
\begin{eqnarray}\label{e5.70}
q_a(x) \equiv \overline{G} Q_a = -2 \int\limits_x^{\infty} \chi^{-2}(y) dy
 \int\limits_y^{\infty} \chi^2(z)Q_a(z) dz
\end{eqnarray}
and for $0 \le x<1$,
\begin{eqnarray}\label{e5.71}
q_a(x) \equiv q_a(1) + 2 \int\limits_x^1 \chi^{-2}(y) dy
 \int\limits_0^y \chi^2(z)Q_a(z) dz,
\end{eqnarray}
where the subscript $a=1,~2,~3$. Because of (\ref{e5.45})-(\ref{e5.46}) and
(\ref{e5.66}), we see that
\begin{eqnarray}\label{e5.72}
\kappa^n_j(x) = q_1(x) + q_2(x) + q_3(x)
\end{eqnarray}
for all $x$.

  On account of (\ref{e5.43}) and (\ref{e5.63}),
\begin{eqnarray}\label{e5.73}
\Delta_{n+j} - \Delta_n &=& \frac{\{\nu k_{n-1}\}}{ \{ k_{n-1}\}}
- \frac{\{\nu (k_{n-1}+\kappa^{n-1}_j)\}} {\{ k_{n+j-1}\}}\nonumber\\
&=& \frac{\{\nu k_{n-1}\}\{k_{n-1}+\kappa^{n-1}_j\}
    -\{\nu (k_{n-1}+\kappa^{n-1}_j)\}\{ k_{n-1}\}}
   {\{ k_{n+j-1}\} \{ k_{n-1}\}}\nonumber\\
&=& - \frac{\Delta_n \{\kappa_j^{n-1}\}}{\{ k_{n+j-1}\}}
    - \frac{\{\nu \kappa_j^{n-1}\}}{\{ k_{n+j-1}\}}~.
\end{eqnarray}
From (\ref{e5.48})-(\ref{e5.49}), we have $ k_{n+j-1}(x) \ge {\cal K}$;
consequently, the absolute value of the above expression satisfies
\begin{eqnarray}
|\Delta_{n+j} - \Delta_n| < \frac{1}{{\cal K}\{ 1 \}} \Big( \Delta_n
|\{\kappa_j^{n-1}\}| + |\{\nu \kappa_j^{n-1}\}|\Big).\nonumber
\end{eqnarray}

  By using the definitions (\ref{e5.64}) for $\overline{\kappa}~^n_j $ and (\ref{e5.43})
for $\Delta_1$ (when $n=1$), together with (\ref{e5.50}), we derive
\begin{eqnarray}\label{e5.74}
|\Delta_{n+j} - \Delta_n| &<& \overline{\kappa}~_j^{n-1}
\Big( \Delta_n + \frac{|\{\nu \}|}{\{ 1 \}}\Big) \Big/ {\cal K}\nonumber\\
&=& \overline{\kappa}~_j^{n-1}(\Delta_n + \Delta_1)\Big/ {\cal K}  \nonumber\\
 &\leq & 2\Delta_1  \overline{\kappa}~_j^{n-1} \Big/ {\cal K} ~.
\end{eqnarray}

  Since $0<k_{n+j-1}(x) < 1$, from (\ref{e5.67}), it follows that
\begin{eqnarray}\label{e5.75}
|Q_1|< 2\Delta_1  \overline{\kappa}~_j^{n-1} \Big/ {\cal K},
\end{eqnarray}
and, because of (\ref{e5.70})-(\ref{e5.71}),
\begin{eqnarray}\label{e5.76}
|q_1(x)|< 2{\cal J} \Delta_1  \overline{\kappa}~_j^{n-1} \Big/ {\cal K}
\end{eqnarray}
where
\begin{eqnarray}\label{e5.77}
{\cal J} \equiv {\cal J}^+ + {\cal J}^-
\end{eqnarray}
with ${\cal J}^+$ defined by (\ref{e5.59}) and
\begin{eqnarray}\label{e5.78}
 {\cal J}^- \equiv 2 \int\limits_0^1  \chi^{-2}(y)dy \int\limits_0^y
 \chi^2(z)dz~.
\end{eqnarray}

  Next,  we examine $Q_2$ given by (5.68). Note that $0< \Delta_n <
\Delta_1$, according to (\ref{e5.50}). Therefore, by using
(\ref{e5.63})-(\ref{e5.64}), we fined
\begin{eqnarray}\label{e5.79}
|Q_2|< \Delta_1  \overline{\kappa}~_j^{n-1},
\end{eqnarray}
which, on account of (\ref{e5.70})-(\ref{e5.71}), leads to
\begin{eqnarray}\label{e5.80}
|q_2(x)| < {\cal J} \Delta_1  \overline{\kappa}~_j^{n-1}~.
\end{eqnarray}
For $Q_3$, multiply its definition (5.69) by $\chi$. Using
(\ref{e5.10}), we have
\begin{eqnarray}\label{e5.81}
\chi Q_3 = - \gamma \psi_+ (k_{n+j-1} - k_{n-1})~.
\end{eqnarray}
Both $\gamma$ and $\psi_+$ are positive. Thus,
\begin{eqnarray}\label{e5.82}
|\chi Q_3|<\gamma \psi_+  \overline{\kappa}~_j^{n-1}~.
\end{eqnarray}
Combining this expression with (\ref{e5.70})-(\ref{e5.71}), we
find for $x \geq 1$
\begin{eqnarray}\label{e5.83}
|q_3(x)|< 2\gamma  \overline{\kappa}~_j^{n-1} \int\limits_x^{\infty} \chi^{-2}(y)dy
\int\limits_y^{\infty} \chi(z) \psi_+(z)dz
\end{eqnarray}
and for $0\leq x<1$
\begin{eqnarray}\label{e5.84}
|q_3(x)|< |q_3(1)| + 2\gamma  \overline{\kappa}~_j^{n-1} \int\limits_x^1 \chi^{-2}(y)dy
\int\limits_0^y \chi(z) \psi_+(z)dz
\end{eqnarray}

  As will be established in (C.53) of Appendix C, for $x>1$
\begin{eqnarray}\label{e5.85}
\psi_-(x)< \gamma J_+ \psi_+(x)
\end{eqnarray}
where $J_+~=~O(\frac{1}{g}\ln \sqrt{g})$ is bounded by
(\ref{e4.29}); therefore
\begin{eqnarray}
\chi(x)=\psi_+(x) - \psi_-(x)>(1-\gamma J_+)\psi_+(x);\nonumber
\end{eqnarray}
i.e.,
\begin{eqnarray}\label{e5.86}
\chi^{-2}(x)<(1-\gamma J_+)^{-2}\psi_+^{-2}(x).
\end{eqnarray}
This together with $\chi(x)>\psi_+(x)$ give, for $x>1$,
\begin{eqnarray}\label{e5.87}
|q_3(x)|< 2\gamma  \overline{\kappa}~_j^{n-1}(1-\gamma J_+)^{-2}
\int\limits_1^{\infty}\psi_+^{-2}(y)dy \int\limits_y^{\infty}
\psi_+^2(z)dz~.
\end{eqnarray}

  Note that for $y<z$
\begin{eqnarray}
\psi_+^{-2}(y)\psi_+^2(z)= \frac{f_+^2(z)}{f_+^2(y)} \phi_+^{-2}(y)
\phi_+^2(z)<  \phi_+^{-2}(y) \phi_+^2(z).\nonumber
\end{eqnarray}
Since, according to (\ref{e4.27}), $2\int\limits_1^{\infty}\phi_+^{-2}(y)dy
\int\limits_y^{\infty}\phi_+^2(z)dz = J_+$, we derive for $x>1$
\begin{eqnarray}\label{e5.88}
|q_3(x)|<\gamma J_+ (1- \gamma J_+)^{-2} \overline{\kappa}~_j^{n-1}.
\end{eqnarray}

  From (\ref{e5.84}) and the definition of $~{\cal I}$ given by
(\ref{e5.58}), it follows that, for $x<1$,
\begin{eqnarray}\label{e5.89}
|q_3(x)|<\gamma~ [ {\cal I} +  J_+ (1- \gamma J_+)^{-2}]~
\overline{\kappa}~_j^{n-1}~.
\end{eqnarray}

  Combining together (\ref{e5.76}),(\ref{e5.80}),(\ref{e5.88}) and
(\ref{e5.89}), we obtain
\begin{eqnarray}\label{e5.90}
\frac{\overline{\kappa}~_j^n}{\overline{\kappa}~_j^{n-1}} < \gamma
\Lambda \equiv r
\end{eqnarray}
where
\begin{eqnarray}\label{e5.91}
\Lambda = {\cal I} + J_+ (1-\gamma J_+)^{-2} +
{\cal J}\frac{\Delta_1}{\gamma}(1+\frac{2}{{\cal K}}),
\end{eqnarray}
confirming (\ref{e5.65}). The bound of $J_+ = O\Big(\frac{1}{g}\ln \sqrt{g}\Big)$
is given by (\ref{e4.29}). Explicit bounds for
\begin{eqnarray}
{\cal I}=O(g^{-\frac{4}{3}}),~~~~~~{\cal J}=O\Big(\frac{1}{g}\ln \sqrt{g}\Big)
+ O(g^{-\frac{4}{3}})~,\nonumber
\end{eqnarray}
\begin{eqnarray}\label{e5.92}
\frac{\Delta_1}{\gamma} \cong 1  {\sf ~~and~~}\gamma \cong 4g \sqrt{\frac{2\pi}{g}}
e^{-\frac{4}{3}g}
\end{eqnarray}
are given by (C.86)-(C.87), (C.109), (C.118) and (C.1)-(C.8)
of Appendix C. The bound for ${\cal K}$ is given by (\ref{e5.60}).
The lemma is then established.

  From (\ref{e5.48}), we see that when $n=1$, $ \overline{\kappa}~_j^1$
satisfies
\begin{eqnarray}\label{e5.93}
\overline{\kappa}~_j^1= \max |k_{1+j}(x)-k_1(x)|<2~.
\end{eqnarray}
Thus, (\ref{e5.90}) gives
\begin{eqnarray}\label{e5.94}
\overline{\kappa}~_j^n~<~r\overline{\kappa}~_j^{n-1}~<~
r^2 \overline{\kappa}~_j^{n-2}~<~\cdots~<~r^{n-1}
\overline{\kappa}~_j^1~<~2r^{n-1}~.
\end{eqnarray}
Analogous to (B.24)-(B.25), $\lim\limits_{n\rightarrow \infty}k_n(x)$
exists for all $x$ and likewise  $\lim\limits_{n\rightarrow \infty}
\Delta_n$ exists. Theorem 5.4 is proved.

\newpage

\section*{\bf Acknowledgment}
\setcounter{section}{8}
\setcounter{equation}{0}

One of us (R. F.) wishes to thank Dr. J. Zinn-Justin for an illustrating discussion
of his method.

\section*{\bf References}
\setcounter{section}{9}
\setcounter{equation}{0}

\noindent
1. A. M. Polyakov, Nucl.Phys. B121 (1977), 429 \\
~2. G. 't Hooft, in: The why's of subnuclear physics. Erice, 1977, ed. A.Zichichi

\noindent ~~~~~~~~~~        (Plenum, New York, 1977)\\
~3. E. Brezin, G. Parisi and J. Zinn-Justin, Phys.Rev. D16 (1977), 408\\
~4. J. Zinn-Justin, J.Math.Phys. 22 (1981), 511 \\
~5. J. Zinn-Justin, Nucl.Phys. B192 (1981), 125 \\
~6. J. Zinn-Justin, in: Recent advances in field theory and statistical mechanics,

\noindent ~~~~~~~~~eds. J.-D. Zuber and R. Stora (Les Houches, session XXXIX, 1982) \\
~7. J. Zinn-Justin, Private Communication \\
~8. Sidney Coleman, in: Aspects of Symmetry, Press Syndicate of the University of

\noindent ~~~~~~~~~Cambridge 1987  \\
~9. E. Shuryak, Nucl.Phys. B302 (1988), 621 \\
10. S. V. Faleev and P. G. Silvestrov, Phys.Lett. A197 (1995), 372 \\
11. R. Friedberg, T. D. Lee and W. Q. Zhao, Ann.Phys. 288 (2001), 52\\
12. R. Friedberg, T. D. Lee and W. Q. Zhao, IL Nuovo Cimento A112 (1999), 1195\\
13. P. M. Morse and H. Feshbach, Methods of Theoretical Physics Part {\rm I},

\noindent ~~~~~~~~~McGraw-Hill Co. (New York), 1953.

\newpage

\section*{\bf  Appendix A }
\setcounter{section}{7}
\setcounter{equation}{0}

In this Appendix we give the proof of Theorem 3.3 of Section 3, together with some related
inequalities. The bounds for ${\cal E}_1$, $J$ and $I$ of (\ref{e2.11})-
(\ref{e2.15}) will be given below in Sections (i), (ii) and (iii).

\noindent
(i) We shall first establish the following three statements (A.1), (A.2)-(A.3) and
(A.4)-(A.5).
$$
 \int\limits_0^{\infty} \phi^2(x) \hat{g}(x) dx <2(2g+3)e^{-\frac{4}{3}g}.\eqno (A.1)
$$
Write
$$
\frac{\int\limits_0^{\infty}\phi_+^2(x)u(x)dx}{\int\limits_0^{\infty}\phi_+^2(x)dx}
=\frac{1}{4} + \frac{9}{2^6}~\frac{1}{g} + a_1; \eqno (A.2)
$$
then
$$
a_1<\frac{311}{2^6g^2}. \eqno (A.3)
$$
Furthermore, if we neglect $O(e^{-\frac{4}{3}g})$, $a_1$ is positive. According
to ({\ref{e2.11}), the first iterated energy ${\cal E}_1$ can be written as
$$
{\cal E}_1 = \frac{1}{4} + \frac{9}{2^6}~\frac{1}{g} + \delta_1. \eqno (A.4)
$$
As we shall see,
$$
\delta_1 = a_1 + O(e^{-\frac{4}{3}g}) . \eqno (A.5)
$$
\underline{Proof}~~ From (2.6) and (2.7), we have
$$
\frac{\phi_-}{\phi_+} = e^{2gS_0-\frac{4}{3}g}. \eqno (A.6)
$$
For $0<x<1$, because of (\ref{e1.8}) and (\ref{e1.15}),
$$
\hat{g}=2g\frac{g-1}{g+1} \frac{\phi_-}{\phi_+} \bigg/ \big(1+(\frac{g-1}{g+1})
\frac{\phi_-}{\phi_+}\big) \eqno (A.7)
$$
and
$$
\phi^2=\frac{4}{(1+x)^2}e^{-2gS_0}\big(1+(\frac{g-1}{g+1})
\frac{\phi_-}{\phi_+}\big)^2; \eqno (A.8)
$$
therefore,
$$
\phi^2 \hat{g} = \frac{8 g}{(1+x)^2}(\frac{g-1}{g+1})e^{-\frac{4}{3}g}\big(1+(\frac{g-1}{g+1})
e^{2gS_0-\frac{4}{3}g}\big). \eqno (A.9)
$$

Since $\hat{g} = 0$ for $x>1$, the left side of (A.1) equals
$$
\int\limits_0^1 \phi(x)^2 \hat{g}(x) dx = 4 g(\frac{g-1}{g+1})e^{-\frac{4}{3}g}(1+r),
\eqno (A.10)
$$
where
$$
r = 2(\frac{g-1}{g+1})\int\limits_0^1 \frac{1}{(1+x)^2}
e^{2gS_0-\frac{4}{3}g}dx. \eqno (A.11)
$$
On account of (\ref{e1.4}),
$$
e^{2gS_0-\frac{4}{3}g} =e^{-2gx+\frac{2}{3}gx^3}. \eqno (A.12)
$$
For $0<x<1$, we have
$$
e^{-2gx}<e^{2gS_0-\frac{4}{3}g}<e^{ -\frac{4}{3}gx} \eqno (A.13)
$$
and
$$
\frac{1}{4} < \frac{1}{(1+x)^2}<1. \eqno (A.14)
$$
Hence,
$$
\frac{1}{8g}(1-e^{-2g})<\int\limits_0^1 \frac{1}{(1+x)^2}e^{2gS_0-\frac{4}{3}g}dx
<\frac{3}{4g}(1-e^{-\frac{4}{3}g}). \eqno (A.15)
$$
Combining (A.10)-(A.11) and (A.15), we derive
$$
\int\limits_0^{\infty} \phi^2(x) \hat{g}(x) dx < 4 g(\frac{g-1}{g+1})e^{-\frac{4}{3}g}
\{1+\frac{3}{2g}(\frac{g-1}{g+1})(1-e^{-\frac{4}{3}g})\},
\eqno (A.16)
$$
and
$$
\int\limits_0^{\infty} \phi^2(x) \hat{g}(x) dx > 4 g(\frac{g-1}{g+1})e^{-\frac{4}{3}g}
\{1+\frac{1}{4g}(\frac{g-1}{g+1})(1-e^{-2g})\}.
\eqno (A.17)
$$
The inequality (A.16) implies (A.1).

Next we turn to (A.2). From (\ref{e1.11}), we have
$$
u(x) - \frac{1}{4} = -(x-1)\bigg( \frac{1}{2(1+x)^2} +  \frac{1}{2^2(1+x)}\bigg).
 \eqno (A.18)
$$
By partial integration and using (2.6), we obtain
\begin{eqnarray}
\int \phi_+^2(x) (u(x)-\frac{1}{4}) dx &=&
-e^{-2gS_0}\bigg\{\frac{1}{2gS_0'}\frac{4}{(1+x)^2}(u(x)-\frac{1}{4})
\bigg\}\nonumber\\
&&+ \int e^{-2gS_0}
\bigg\{ \frac{1}{2gS_0'}\frac{4}{(1+x)^2}(u(x)-\frac{1}{4}) \bigg\}' dx,
~~~~~~~~~~~~~~(A.19)\nonumber
\end{eqnarray}
which, on account of (A.18) and $S_0'=x^2-1$, leads to
$$
\int\limits_0^{\infty} \phi_+^2(x) \big(u(x)-\frac{1}{4}\big) dx =
- e^{-\frac{4}{3}g} \frac{3}{2g}
+ \frac{1}{g} \int\limits_0^{\infty} e^{-2gS_0}
\bigg(\frac{5}{(1+x)^6} + \frac{2}{(1+x)^5}\bigg) dx.
\eqno (A.20)
$$
Next, write
$$
\int\limits_0^{\infty} \phi_+^2 \frac{9}{2^6 g} dx =
\frac{1}{g} \int\limits_0^{\infty} e^{-2gS_0}
\frac{1}{(1+x)^2}\Big(\frac{5}{2^4} + \frac{2}{2^3}\Big) dx.
\eqno (A.21)
$$
Taking the difference (A.20) minus (A.21) and doing another partial integration, we
derive
(A.2), in which
\begin{eqnarray}
a_1&=&\frac{1}{\int\limits_0^{\infty} \phi_+^2(x) dx}\bigg\{-e^{-\frac{4}{3}g}
\frac{1}{2g}(3+\frac{103}{16}~\frac{1}{g})\nonumber\\
&&+ \int\limits_0^{\infty} e^{-2gS_0}\frac{1}{4g^2}
\frac{1}{(1+x)^5}\bigg(\frac{35}{(1+x)^3} + \frac{54}{2(1+x)^2} + \frac{45}{2^2(1+x)}
+ \frac{36}{2^3}\bigg) dx \bigg\},
~~~~~ (A.22)\nonumber
\end{eqnarray}
in which the integral in the numerator is less than
$$
\int\limits_0^{\infty} e^{-2gS_0}(\frac{2}{1+x})^2 \frac{1}{16g^2}(35
+ \frac{54}{2}+ \frac{45}{2^2}+ \frac{36}{2^3}) dx
= \frac{311}{2^6}~\frac{1}{g^2}\int\limits_0^{\infty}\phi_+^2(x) dx.
\eqno (A.23)
$$
Therefore, (A.3) follows. Furthermore, if we neglect the $O(e^{-\frac{4}{3}g})$ term
in (A.22), $a_1$ is positive.

To establish (A.4)-(A.5), we define
$$
A \equiv \int\limits_0^{\infty}\phi^2 dx - \int\limits_0^{\infty}\phi_+^2 dx
\eqno (A.24)
$$
and
$$
B \equiv \int\limits_0^{\infty}\phi^2 u dx - \int\limits_0^{\infty}\phi_+^2 u dx.
\eqno (A.25)
$$
From (\ref{e1.8}), one can readily verify that
\begin{eqnarray}
A &=& 2 (\frac{g-1}{g+1})e^{-\frac{4}{3}g}
 \bigg\{(1+\frac{1}{2}(\frac{g-1}{g+1})e^{-\frac{4}{3}g})  \int\limits_1^{\infty}\phi_+^2 dx
 \nonumber\\
&& +2 +2(\frac{g-1}{g+1}) \int\limits_0^1 \frac{1}{(1+x)^2} e^{2gS_0-\frac{4}{3}g}
 dx\bigg\}\nonumber
\end{eqnarray}
and
\begin{eqnarray}
B &=& 2 (\frac{g-1}{g+1})e^{-\frac{4}{3}g}
 \bigg\{(1+\frac{1}{2}(\frac{g-1}{g+1})e^{-\frac{4}{3}g})  \int\limits_1^{\infty}\phi_+^2 u dx
 \nonumber\\
&& +\frac{7}{6} +2(\frac{g-1}{g+1}) \int\limits_0^1 \frac{1}{(1+x)^4} e^{2gS_0-\frac{4}{3}g}
 dx\bigg\}~.  \nonumber
\end{eqnarray}
Since for $x>1$, $\phi_+^2(x)<e^{-2g(x-1)^2}$, we derive
$$
\int\limits_1^{\infty}\phi_+^2 dx < \frac{1}{2}\sqrt{\frac{\pi}{2g}}~~~~{\sf and}
~~~~\int\limits_1^{\infty}\phi_+^2 u dx <
\frac{1}{8}\sqrt{\frac{\pi}{2g}}.\eqno(A.26)
$$
By using (A.13)-(A.15), we have
$$
 \int\limits_0^1 \frac{1}{(1+x)^4} e^{2gS_0-\frac{4}{3}g}dx <
 \int\limits_0^1 \frac{1}{(1+x)^2} e^{2gS_0-\frac{4}{3}g}dx<\frac{3}{4g}.\nonumber
$$
It follows then
$$
4(\frac{g-1}{g+1})e^{-\frac{4}{3}g} < A <
4 e^{-\frac{4}{3}g}\bigg\{1+\frac{1}{4}\sqrt{\frac{\pi}{2g}}
(1+\frac{1}{2}e^{-\frac{4}{3}g})+\frac{3}{4g}\bigg\}\eqno(A.27)
$$
and
$$
\frac{7}{3}(\frac{g-1}{g+1})e^{-\frac{4}{3}g} < B <
\frac{7}{3} e^{-\frac{4}{3}g}\bigg\{1+\frac{3}{28}\sqrt{\frac{\pi}{2g}}
(1+\frac{1}{2}e^{-\frac{4}{3}g})+\frac{9}{7g}\bigg\}~.\eqno(A.28)
$$
Therefore,
$$
A = O(e^{-\frac{4}{3}g})~~~~{\sf and}~~~~B = O(e^{-\frac{4}{3}g}) \nonumber
$$
Since
$$
{\cal E}_1 = \frac{\int\limits_0^{\infty}\phi^2 (u + \hat{g})dx}
{\int\limits_0^{\infty}\phi^2 dx}~, \eqno (A.29)
$$
we have
$$
{\cal E}_1 = \frac{\int\limits_0^{\infty}\phi_+^2 u dx + B +
\int\limits_0^{\infty}\phi^2\hat{g}dx}
{\int\limits_0^{\infty}\phi_+^2 dx +A}~, \eqno (A.30)
$$
From (A.1), (A.2), (A.27) and (A.28), we see that (A.5) holds; i.e.,
$$
{\cal E}_1 = \frac{1}{4} + \frac{9}{2^6}~\frac{1}{g} + a_1 + O(e^{-\frac{4}{3}g}).
 \eqno (A.31)
$$
with $a_1 < 311/2^6g^2$. Thus we establish (\ref{e2.11})-(\ref{e2.12}) of Theorem 3.3.

\noindent
(ii) To set a bound on the integral $J$ of (\ref{e2.14}), we first introduce
$$
j(x) \equiv e^{2gS_0(x)}\int\limits_x^{\infty}e^{-2gS_0(z)}dz
 \eqno (A.32)
$$
and then show, for $x>1$,
$$
j(x) < \frac{C}{(1+x)^2}, \eqno (A.33)
$$
where
$$
C = \frac{1}{2g}~\frac{l+1}{l-1} \eqno (A.34)
$$
and
$$
l = \sqrt{1+\sqrt{2/\pi g}}. \eqno (A.35)
$$
Thus,
$$
C = \sqrt{\frac{2 \pi}{g}}(1+\frac{3}{2\sqrt{2 \pi g}} + \cdots) \eqno (A.36)
$$
and, (A.33) gives
$$
\int\limits_1^{\infty}j(x)dx <\frac{1}{2} C. \eqno (A.37)
$$
As we shall see, the same $C$ is also an upper bound for $J$:
$$
J \equiv 2 \int\limits_1^{\infty} \phi^{-2}(y) dy \int\limits_y^{\infty}\phi^2(z) dz
 < C. \eqno (A.38)
$$

Next, we shall improve this bound by proving
\begin{eqnarray}
J &<& (l-1) \sqrt{\frac{\pi}{2 g}}+ \frac{1}{2 g}~ln~\frac{l+1}{l-1}
~~~~~~~~~~~~~~~~~~~~~~~~~~~~~~~~~~~~~~~~~~~~~~~~~~~~~~~~~~~~~~~ (A.39.1)\nonumber\\
  &<& \frac{1}{2 g}~ln~(1+2\sqrt{2 \pi g}) +  \frac{1}{2 g}.
~~~~~~~~~~~~~~~~~~~~~~~~~~~~~~~~~~~~~~~~~~~~~~~~~~~~~~~~~~~~~~~~ (A.39.2)\nonumber
\end{eqnarray}
where $l$ is the same one given above.

\noindent
\underline{Proof}~~~For $x>1$, $S_0(x) = \frac{1}{3}(x-1)^2(x+2)>(x-1)^2$ and
therefore at $x=1$
$$
j(1) < \int\limits_1^{\infty}  e^{-2g(x-1)^2} dx = \frac{1}{2} \sqrt{\frac{\pi}{2g}}.
 \eqno (A.40)
$$
Differentiating (A.32), we find the derivative of $j(x)$ to be
$$
j'(x) = 2g(x^2-1)j(x)-1. \eqno (A.41)
$$
Using (A.35) to define $l$, we see that $l$ satisfies
$$
2g(l^2-1) \frac{1}{2} \sqrt{\frac{\pi}{2g}}-1=0. \eqno (A.42)
$$
It follows then, for $1<x<l$, $j'(x)<0$ and therefore
$$
j(x)<j(1)< \frac{1}{2} \sqrt{\frac{\pi}{2g}}. \eqno (A.43)
$$
Furthermore, partial integrating (A.32), we find
$$
j(x) = \frac{1}{2g(x^2-1)} -  e^{2gS_0(x)}\int\limits_x^{\infty}e^{-2gS_0(z)}
 \frac{z}{g(z^2-1)^2}dz, \nonumber
$$
which gives, for $x>1$,
$$
j(x) < \frac{1}{2g(x^2-1)}. \eqno (A.44)
$$

The constant $C$ in (A.33) is determined by setting
$$
\frac{C}{(l+1)^2} = \frac{1}{2g(l^2-1)}. \eqno (A.45)
$$
For $x>l>1$, since
$$
\frac{l+1}{l-1} > \frac{x+1}{x-1};  \nonumber
$$
we have $x^2-1 > (1+x)^2(l-1)/(l+1)$, and because of (A.44)-(A.45), we obtain
$$
j(x) <  \frac{l+1}{2g(1+x)^2(l-1)} = \frac{C}{(1+x)^2}, \eqno (A.46)
$$
which becomes, at $x=l$,
$$
j(l) < \frac{C}{(1+l)^2} = \frac{1}{2} \sqrt{\frac{\pi}{2g}}, \eqno (A.47)
$$
on account of (A.42) and (A.45). Combining (A.43) with (A.46)-(A.47), we prove
(A.33).

For $z>y>1$,
\begin{eqnarray}
\phi^{-2}(y) \phi^2(z) &=& e^{2gS_0(y)-2gS_0(z)} (\frac{1+y}{1+z})^2\nonumber\\
                       &<& e^{2gS_0(y)-2gS_0(z)}. ~~~~~~~~~~~~~~~~~~~~~~~~~
                       ~~~~~~~~~~~~~~~~~~~~~~~~~~~~~~~~~~~~~~~ (A.48)
                       \nonumber
\end{eqnarray}
Thus, (A.37) implies (A.38).

To improve the bound (A.36), we observe that
\begin{eqnarray}
S_0(z)-S_0(y) &=& (z-y)(\frac{1}{3}(z^2+zy+y^2)-1)\nonumber\\
              &=& (z-y)(\frac{1}{3}(z-y)^2+y(z-y)+y^2-1). \nonumber
\end{eqnarray}
Thus, for $z>y>1$,
$$
S_0(z)-S_0(y) > (z-y)^2 \eqno (A.49)
$$
and also
$$
S_0(z)-S_0(y) > (z-y)(y^2-1). \eqno (A.50)
$$
Next, we separate the integration range in (A.38) for $J$ into two regions: 1.
$1<y<a$, and 2. $y>a$; correspondingly,
$$
J=J_1 + J_2, \eqno (A.51)
$$
where
$$
J_1 = 2 \int\limits_1^a dy \int\limits_y^{\infty} dz \phi^{-2}(y)\phi^2(z) \nonumber
$$
and
$$
J_2 = 2 \int\limits_a^{\infty} dy \int\limits_y^{\infty} dz \phi^{-2}(y)\phi^2(z).
 \eqno (A.52)
$$

From (A.48)-(A.49), we have
$$
J_1 < 2 \int\limits_1^a dy \int\limits_y^{\infty} e^{-2g(z-y)^2} dz
=(a-1)\sqrt{\frac{\pi}{2g}} \eqno (A.53)
$$
and
$$
J_2 < 2 \int\limits_a^{\infty} dy \int\limits_y^{\infty} e^{-2g(z-y)(y^2-1)} dz
=\frac{1}{g}\int\limits_a^{\infty} \frac{dy}{y^2-1}
=\frac{1}{2g}~ln~\frac{a+1}{a-1}~.
 \eqno (A.54)
$$
Thus, for any $a>1$,
$$
J <(a-1)\sqrt{\frac{\pi}{2g}} + \frac{1}{2g}~ln~\frac{a+1}{a-1}~, \eqno (A.55)
$$
which is minimum when
$$
a^2-1 = \sqrt{\frac{2}{\pi g}}. \eqno (A.56)
$$
Comparing the above equation with (A.42), we see that
$$
a = l = \sqrt{1+\sqrt{\frac{2}{\pi g}}} = 1 + \frac{1}{2}\sqrt{\frac{2}{\pi g}}
+\cdots~; \eqno(A.57)
$$
consequently, (A.39.1) is established. By setting $a = 1+ (2 \pi g)^{-\frac{1}{2}}$,
instead of $l$, we obtain the second bound (A.39.2) for $J$.

\noindent
(iii)~~~For any function $F(x)$, define
$$
I(F) \equiv 2  \int\limits_0^1 \phi^{-2}(y) dy \int\limits_0^y \phi^2(z)F(z) dz.
\eqno (A.58)
$$
The integral $I$ of (\ref{e2.13}) can be written as
$$
I = I(w) = I(u) + I(\hat{g}) \eqno (A59)
$$
with $w$, $u$ and $\hat{g}$ given by (\ref{e1.14}), (\ref{e1.11}) and (\ref{e1.15}).
For $I(u)$, we have,
on account of (\ref{e1.8}),
$$
I(u) < 8 \int\limits_0^1 dy  \int\limits_0^y dz  e^{-2g(S_0(z)-S_0(y))}
\frac{(1+y)^2}{(1+z)^4}~.
\eqno (A.60)
$$

Next, divide the integration range into two regions: 1. $0<z<y<b<1$ and 2. $b<y<1$ and
$0<z<y$. In region 1
$$
S_0(z)-S_0(y) = (y-z)(1-\frac{1}{3}(y^2+yz+z^2))>(y-z)(1-\frac{1}{3}(y+y+1))
\eqno (A61)
$$
and in region 2
$$
S_0(z)-S_0(y) >(y-z)(y-\frac{1}{3}(y+z+z)).
\eqno (A62)
$$
In both regions
$$
\frac{(1+y)^2}{(1+z)^4}<2^2. \eqno (A.63)
$$
Thus, (A.60) leads to
\begin{eqnarray*}
I(u) &<& 32 \int\limits_0^b dy  \int\limits_0^y dz ~e^{-\frac{4}{3}g(y-z)(1-y)}\\
     && +32 \int\limits_b^1 dy  \int\limits_0^y dz ~e^{-\frac{4}{3}g(y-z)^2}\\
     &=& 32 \int\limits_0^b dy \frac{3}{4g(1-y)}(1- e^{-\frac{4}{3}g y(1-y)})\\
     && +32 \int\limits_b^1 dy  \int\limits_0^y dz ~e^{-\frac{4}{3}g(y-z)^2}\\
     &<& 32 \bigg\{-\frac{3}{4g}~ln~(1-b)+\frac{1}{4}(1-b)\sqrt{\frac{3\pi}{g}}\bigg\}
\end{eqnarray*}
whose minimum is at
$$
b=1-\sqrt{\frac{3}{\pi g}}. \eqno(A.64)
$$
we obtain
$$
I(u) < \frac{24}{g}(ln~\sqrt{\frac{\pi g}{3}}+1). \eqno (A.65)
$$

To analyze $I(\hat{g})$, we start from (A.9). Since $(1+x)^{-2}<1$, $(g-1)/(g+1)<1$
and for $0<x<1$, $e^{2gS_0-\frac{4}{3}g}<1$, we have
$$
\phi^2(x) \hat{g}(x) < 16g~e^{-\frac{4}{3}g}. \eqno (A.66)
$$
From (2.6)-(\ref{e1.8}), we have
$$
\phi^{-2}(x) e^{-\frac{4}{3}g}<e^{2gS_0-\frac{4}{3}g}, \nonumber
$$
which, on account of (A.13), is $< e^{-\frac{4}{3}gx}$ for $0<x<1$. Thus
\begin{eqnarray*}
I(\hat{g}) &=& 2 \int\limits_0^1 \phi^{-2}(y)dy \int\limits_0^y \phi^2(z)\hat{g}(z)dz\\
         &<& 32g \int\limits_0^1  e^{-\frac{4}{3}gy} y dy < \frac{18}{g}.
         ~~~~~~~~~~~~~~~~~~~~~~~~~~~~~~~~~~~~~~~~~~~~~~~~~~~~~~~~~~~~~~~~~~(A.67)
\end{eqnarray*}
Combining (A.65) and (A.67) we derive
$$
I=I(w)<\frac{24}{g}(ln~\sqrt{\frac{\pi g}{3}} + \frac{7}{4}). \eqno(A.68)
$$

\newpage

\section*{\bf  Appendix B }
\setcounter{section}{8}
\setcounter{equation}{0}

To establish  Theorem 3.6, we introduce
$$
\delta_j^n(x) \equiv f_{n+j}(x)-f_n(x) \eqno (B.1)
$$
and
$$
\Delta_j^n \equiv {\sf maximum~of~} |\delta_j^n(x)| ~~~{\sf over~all~}
~~~x \geq 0.  \eqno(B.2)
$$
As we shall see,

\noindent
\underline{Lemma}~~~ For $g$ sufficiently large,
$$
\frac{\Delta_j^n}{\Delta_j^{n-1}} \leq R << 1,  \eqno (B.3)
$$
where $R$ is independent of $n$ and $j$.

\noindent
\underline{Proof}~~~  From (\ref{e1.27}) we can write
\begin{eqnarray*}
~~~~~~~~~~~~~~~~~\delta_j^n &=& \overline{D}\big( (w-{\cal E}_{n+j})f_{n+j-1} -
(w-{\cal E}_n)f_{n-1}\big)\\
           &\equiv & \overline{D} (L_1+L_2+L_3),~~~~~~~~~~~~~~~~~~~~~~
           ~~~~~~~~~~~~~~~~~~~~~~~~~~~~~~~~~~~(B.4)
\end{eqnarray*}
where
\begin{eqnarray*}
~~~~~~~~~~~~~~~~L_1 &=& ({\cal E}_n-{\cal E}_{n+j})f_{n+j-1} ~~~~~~~~~~~~~~~~~~~~~
~~~~~~~~~~~~~~~~~~~~~~~~~~~~~~~~~~~~~(B.5)\\
L_2 &=& (\frac{1}{4}-{\cal E}_n)(f_{n+j-1}-f_{n-1})~~~~~~~~~~~~~~~
~~~~~~~~~~~~~~~~~~~~~~~~~~~~~~~~~~~(B.6) \\
L_3 &=& (w-\frac{1}{4})(f_{n+j-1}-f_{n-1}).~~~~~~~~~~~~~~~~~~~~~~~~
~~~~~~~~~~~~~~~~~~~~~~~~~~(B.7)
\end{eqnarray*}
Define for $x\geq1$
$$
\lambda_a(x)\equiv \overline{D} L_a=
-2\int\limits_x^{\infty} \phi^{-2}(y) dy \int\limits_y^{\infty} \phi^2(z)
L_a(z)dz \eqno  (B.8)
$$
and for $ 0<x<1$
$$
\lambda_a(x)=\lambda_a(1)+
2\int\limits_x^1 \phi^{-2}(y) dy \int\limits_0^y \phi^2(z)
L_a(z)dz, \eqno  (B.9)
$$
where the subscript $a=1,~2,~3$. Because of (\ref{e1.35})-(\ref{e1.37}) and (B.4),
we have
$$
\delta_j^n(x)=\lambda_1(x)+\lambda_2(x)+\lambda_3(x)                 \eqno (B.10)
$$
for all $x$.

To set a bound for $\lambda_1(x)$, we observe that, on  account of (\ref{e1.28})
and (B.1),
\begin{eqnarray*}
~~~~~~~~~~~~~{\cal E}_n-{\cal E}_{n+j}&=&
\frac{[wf_{n-1}]}{[f_{n-1}]}-
\frac{[w(f_{n-1}+\delta_j^{n-1})]}{[f_{n+j-1}]}\\
&=&\frac{[wf_{n-1}][f_{n-1}+\delta_j^{n-1}]
-[w(f_{n-1}+\delta_j^{n-1})][f_{n-1}]}{[f_{n-1}][f_{n+j-1}]}\\
&=&\frac{{\cal E}_n}{[f_{n+j-1}]}[\delta_j^{n-1}]
-\frac{[w\delta_j^{n-1}]}{[f_{n+j-1}]}.~~~~~~~~~~~~~~~~~~~~~~~
~~~~~~~~~~~~~~~~(B.11)
\end{eqnarray*}
From (\ref{e2.2}), we have $[f_{n+j-1}]>[1]$ and therefore the absolute value of
(B.11) satisfies
$$
|{\cal E}_n-{\cal E}_{n+j}|<\frac{1}{[1]}
({\cal E}_n~|[\delta_j^{n-1}]|~+~|[w\delta_j^{n-1}]|) \nonumber
$$
By using (B.2) and (3.31)  we  find
\begin{eqnarray*}
~~~~~~~~~~~~~~|{\cal E}_n-{\cal E}_{n+j}|&<&\Delta_j^{n-1}({\cal E}_n+\frac{[w]}{[1]})\\
&=&\Delta_j^{n-1}({\cal E}_n+{\cal E}_1)\leq \Delta_j^{n-1}~2K,
~~~~~~~~~~~~~~~~~~~~~~~~~~~~~~~~~(B.12)
\end{eqnarray*}
where $K$ is given by (\ref{e2.27}). Since $1<f_{n+j-1}(x)<f_{n+j-1}(0)$,
from (B.5) it follows that
$$
|L_1|<|{\cal E}_n-{\cal E}_{n+j}|~f_{n+j-1}(0).                               \eqno(B.13)
$$
On account of (B.8)-(B.9) and  (B.12), for $x \ge 1$
$$
|\lambda_1(x)|~<~2~f_{n+j-1}(0)~K \Delta_j^{n-1}2 \int_1^{\infty}\phi^{-2}(y)dy
\int_y^{\infty}\phi^2(z) dz = 2~f_{n+j-1}(0)~K~\Delta_j^{n-1}~J,           \eqno  (B.14)
$$
where according to (A.39.2),
$$
J~<~\frac{1}{2g}\ln(e+2e\sqrt{2\pi g});
$$
for $x<1$
$$
|\lambda_1(x)|~<~|\lambda_1(1)| + 2 f_{n+j-1}(0) K \Delta_j^{n-1}
2 \int_0^1\phi^{-2}(y)dy \int_0^y \phi^2(z) dz
$$
$$
~~~~~~= ~|\lambda_1(1)| + 2~f_{n+j-1}(0)~K~\Delta_j^{n-1} I(1),
                                                                   \eqno(B.15)
$$
where, in accordance with (A.58),
$$
I(1) \equiv 2 \int_0^1\phi^{-2}(y)dy \int_0^y \phi^2(z) dz
$$
$$
~~~~~~~~< 8 \int_0^1 dy \int_0^y dz~ e^{-2g(S_0(z)-S_0(y))}\frac{(1+y)^2}{(1+z)^2}
$$
$$
~~~~~~~~< 32 \int_0^1 dy \int_0^y dz~ e^{-2g(S_0(z)-S_0(y))}
        < \frac{24}{g}\ln\Big( \sqrt{\frac{\pi g}{3}}+1\Big),          \eqno(B.16)
$$
which is, on account of (A.61)-(A.64), the same upper bound for $I(u)$ given by (A.65).
Combining (B.14) with (B.15), we see that at all $x$,
$$
|\lambda_1(x)| < 2 ~f_{n+j-1}(0)~K(J+I(1))\Delta_j^{n-1}.              \eqno(B.17)
$$

Next, we examine $L_2$ and $L_3$ of (B.6)-(B.7). Because of (3.6),
(A.4) and (3.31), $\frac{1}{4}<{\cal E}_n<K$,
$$
|L_2|~<~({\cal E}_n-\frac{1}{4})~|\delta_j^{n-1}|
~<~(K-\frac{1}{4})\Delta_j^{n-1}. \nonumber
$$
Thus, using (B.8)-(B.9) and by following the same steps leading from (B.13) to (B.17),
we obtain
$$
|\lambda_2(x)| <~(K-\frac{1}{4})(J+I(1))\Delta_j^{n-1}.               \eqno(B.18)
$$
Likewise,
$$
|L_3(x)|< |w-\frac{1}{4}|\Delta_j^{n-1}.
$$
Since, from (\ref{e1.11}) and (\ref{e1.14})-(\ref{e1.15}),
for $x>1$, $w(x)=u(x)<\frac{1}{4}$
and for $0<x<1$, $w(x)>\frac{1}{4}$, we have
\begin{eqnarray*}
~~~~~~~~~~~~~~~~~~~|\lambda_3(x)|&<&\Delta_j^{n-1}\bigg\{
2\int\limits_1^{\infty} \phi^{-2}(y) dy \int\limits_y^{\infty} \phi^2(z)
(\frac{1}{4}-u(z))dz \\
&&+2\int\limits_0^1 \phi^{-2}(y) dy \int\limits_0^y \phi^2(z)
(w(z)-\frac{1}{4})dz \bigg\}\\
&<&\Delta_j^{n-1}\{\frac{1}{4}~J+I\}.~~~~~~~~~~~~~~~~~~~~~~~~
~~~~~~~~~~~~~~~~~~~~~~~~ (B.19)
\end{eqnarray*}
Combining (B.17)-(B.19), we derive
$$
\frac{\Delta_j^n}{\Delta_j^{n-1}}~<~\{2f_{n+j-1}(0)~K+(K-\frac{1}{4})\}
(J+I(1))+\frac{1}{4}~J+I.                                             \eqno(B.20)
$$
From (A.68) we have
$$
I<\frac{24}{g}\ln\Big( \sqrt{\frac{\pi g}{3}}+\frac{7}{4}\Big).
$$

Using (\ref{e2.17})-(3.18) and (\ref{e2.26}), we find for any $m \geq 1$,
$$
f_m(0)<\frac{1}{1-I}~\frac{1}{1-JK}~=~\frac{K}{{\cal E}_1};          \eqno(B.21)
$$
therefore
$$
\frac{\Delta_j^n}{\Delta_j^{n-1}}~<~R~=~
\Big(\frac{2K^2}{{\cal E}_1} +K-\frac{1}{4}\Big)(J+I(1))+\frac{1}{4}J+I. \eqno(B.22)
$$
Since for $g$ sufficiently large,
$J$, $I(1)$ and $I$ are all $O(\frac{ln~\sqrt{g}}{g})<<1$, so is $R<<1$.
The Lemma is proved.

From (\ref{e2.2}) and (B.21), when $n=1$, $\Delta_j^1$ satisfies
$$
\Delta_j^1 \equiv {\sf max~} |f_{1+j}(x)-f_1(x)|<f_{1+j}(0)+f_1(0)
<2K/{\cal E}_1.                                                       \eqno(B.23)
$$
Thus, (B.22) implies
$$
\Delta_j^n<R\Delta_j^{n-1}<R^2\Delta_j^{n-2}<\cdots<R^{n-1}\Delta_j^1<
2R^{n-1}K/{\cal E}_1.                                                 \eqno(B.24)
$$
Consequently, given any $\epsilon>0$, there exists an $N$ such that
$(2R^{N-1}K/{\cal E}_1)<\epsilon$, and therefore
$$
\Delta_j^N = {\sf max~} |f_{N+j}(x)-f_N(x)|<\epsilon.         \eqno(B.25)
$$
for all $j$.  Hence, by using Cauchy's test of convergence we establish
the existence of $\lim\limits_{n\rightarrow\infty} f_n(x)$ for  all $x$.
From (\ref{e1.28}), it follows that
$\lim\limits_{n\rightarrow\infty}{\cal E}_n$ also exists. Theorem 3.6
is  then  proved.

\newpage

\section*{\bf  Appendix C }
\setcounter{section}{11}
\setcounter{equation}{0}

  This appendix is divided into several sections. The proof of Theorem 5.2 is given
in Section C.1. In order to set the bound for ${\cal K}$, defined in (\ref{e5.60}),
we need first the bounds for several other relevant functions and integrals. The
bounds for $\psi_-$,$~\psi_-'$ and $\chi'$ are given in Section C.2, the bound for
${\cal I}$ is in Section C.3, the bounds for ${\cal J}^+$ and ${\cal J}^-$ in
Section C.4 and the bound for $\Delta_1$ in Section C.5.
\\

\noindent
C.1.  Proof of Theorem 5.2

  As we shell see, the bounds $\alpha_{\gamma}$ and $\beta_{\gamma}$ in
(\ref{e5.19})-(\ref{e5.20})
$$
\gamma < 4g\sqrt{\frac{2g}{\pi}} e^{-\frac{4}{3}g}( 1 + \alpha_{\gamma}) \eqno (C.1)
$$
$$
\gamma > 4g\sqrt{\frac{2g}{\pi}} e^{-\frac{4}{3}g}( 1 - \beta_{\gamma}) \eqno (C.2)
$$
can be written as
$$
1 + \alpha_{\gamma} = (1-L)^{-2} (1 - e^{-\frac{2}{3}g})^{-1}
( 1 - \frac{3}{2\sqrt{2\pi g}} e^{-\frac{8}{9}g})^{-1} \eqno (C.3)
$$
where $L = O(\frac{1}{g}\ln\sqrt{g})$ is bounded by (\ref{e4.45}), and
$$
1 - \beta_{\gamma} = [(1+\alpha_M)(1+\alpha_N) +
4g\sqrt{\frac{2g}{\pi}} e^{-\frac{4}{3}g}J_{+} ]^{-1} \eqno (C.4)
$$
with $J_{+}$ given by (\ref{e4.27}) and (\ref{e4.29}) and $\alpha_M$, $\alpha_N$
are in turn bounds for
$$
M \equiv \int\limits^{1}_0 \phi^{-2}_{+}(x)dx
< \frac{1}{8g} e^{\frac{4}{3}g} (1+\alpha_M) \eqno (C.5)
$$
and
$$
N \equiv \int\limits^{\infty}_0 \phi^2_{+}(x)dx
< \sqrt{\frac{\pi}{2g}} (1+\alpha_N) \eqno (C.6)
$$
which are given by
$$
1 + \alpha_M = (1 - e^{-2\sqrt{g}}) e^{\frac{2}{3\sqrt{g}}} (1 + \frac{1}{\sqrt{g}})^2
 + 6g(e^{-\frac{4}{3}\sqrt{g}} - e^{-\frac{4}{3}g}) \eqno (C.7)
$$
and
$$
1 + \alpha_N = (1-\frac{3}{4g^\frac{1}{3}})^{-2}(1+ \frac{5}{48g}) +
(1+e^{-\frac{9}{2}})\sqrt{ \frac{2}{\pi}}~\frac{1}{ g^{1/6}} e^{-3g^{\frac{1}{3}}} +
\frac{1}{2\sqrt{2\pi g}}e^{-2g}                                             \eqno (C.8)
$$
Therefore, for $g$ sufficiently large
$$
\alpha_M = O(g^{-\frac{1}{2}}),~~~~\alpha_N = O(g^{-\frac{1}{3}}) \nonumber
$$
$$
\alpha_{\gamma} = O(\frac{1}{g} \ln\sqrt{g})~~~{\sf  and}~~~
\beta_{\gamma} = O(g^{-\frac{1}{3}}). \eqno (C.9)
$$
From these expressions, we anticipate that some of  their derivations can be
somewhat tedious.

From (\ref{e5.8}),
$$
\gamma^{-1} = 2 \int\limits^{\infty}_0 \psi_+^{-2}(y)dy
\int\limits^{\infty}_y\psi_+^2(z)dz.  \eqno (C.10)
$$
Take any $l$ between $0$ and $1$. We have
$$
\gamma^{-1} > 2 \int\limits^{1-l}_0 \psi_+^{-2}(y)dy
\int\limits^{1+l}_{1-l}\psi_+^2(z)dz.  \eqno (C.11)
$$
By using (\ref{e4.5}), (\ref{e4.9}) and (\ref{e4.43}), we see that
$$
\psi_+^{-2}(y)\psi_+^2(z)=\frac{f_+^2(z)}{f_+^2(y)}~\frac{(1+y)^2}{(1+z)^2}
e^{2g(S_0(y)-S_0(z))}
> \frac{e^{2g(S_0(y)-S_0(z))}}{f_+^2(0)(1+z)^2}~. \eqno (C.12)
$$
Since $S_0(y) = \frac{1}{3}y^3 -y +\frac{2}{3}>-y+\frac{2}{3}$,
$$
\int\limits^{1-l}_0 e^{2g S_0(y)} dy > e^{\frac{4}{3}g}\int\limits^{1-l}_0 e^{-2gy}dy
= \frac{1}{2g}e^{\frac{4}{3}g}(1-e^{-2g(1-l)}). \eqno (C.13)
$$

Next, set $z= 1 \mp \xi$ and write $S_0(z) = \xi^2 \mp \frac{1}{3}\xi^3$. We find
$$
\int\limits^{1+l}_{1-l}\frac{e^{-2gS_0(z)}}{(1+z)^2}~dz =
\int\limits^l_0 \frac{2(4+\xi^2)\cosh (\frac{2}{3}g\xi^3) + 8 \xi \sinh (\frac{2}{3}g\xi^3)}
{(4-\xi^2)^2}~e^{-2g\xi^2}d\xi\nonumber
$$
$$
> \int\limits^l_0 \frac{1}{2}e^{-2g\xi^2}d\xi >
\frac{1}{4}\sqrt{\frac{\pi}{2g}}
(1-\frac{1}{\sqrt{2\pi g}}~\frac{1}{l}e^{-2gl^2}). \eqno (C.14)
$$
Combining (C.11)-(C.14) and using (\ref{e4.44}), we derive
$$
\gamma^{-1}>\frac{(1-L)^2}{4g}e^{ \frac{4}{3}g}\sqrt{\frac{\pi}{2g}}(1-e^{-2g(1-l)})
(1-\frac{1}{\sqrt{2\pi g}}~\frac{1}{l}e^{-2gl^2}), \nonumber
$$
which, for $l=\frac{2}{3}$, gives the upper bound (C.1) and (C.3) for $\gamma$.

To derive the lower bound for $\gamma$, we observe that for $ y<z$, in place of
(C.12),
$$
\psi^{-2}_{+}(y) \psi^2_{+}(z) = \frac{f^2_{+}(z)}{f^2_{+}(y)}
\phi^{-2}_{+}(y)\phi^2_{+}(z) < \phi^{-2}_{+}(y)\phi^2_{+}(z).
$$
Substituting the inequality into (C.10) and decomposing the $y$-integration into two
parts: $\int\limits^{\infty}_1 dy$ and  $\int\limits^{1}_0 dy$, we derive
$$
\gamma^{-1} < J_+ + 2 \int\limits^1_0  \phi^{-2}_{+}(y)dy
\int\limits^{\infty}_y  \phi^2_{+}(z)dz
< J_+ + 2 \int\limits^1_0  \phi^{-2}_{+}(y)dy
\int\limits^{\infty}_0  \phi^2_{+}(z)dz      \eqno (C.15)
$$
where, according to (\ref{e4.27}) and (\ref{e4.29}),
$$
J_+ = 2\int\limits_1^{\infty} \phi_+^{-2}(y) dy
 \int\limits_y^{\infty} \phi_+^2(z) dz
< \frac{1}{2g} \ln(e + 2 e \sqrt{2\pi g}).
 \eqno (C.16)
$$
Using the definitions $M$ and $N$ given by (C.5)-(C.6), we see that the inequality
(C.15) can be written as
$$
\gamma^{-1} < J_+ + 2MN ~~~{\sf  or}~~~
\gamma > (J_+ + 2MN)^{-1}.
\eqno (C.17)
$$
\noindent \underline{Lemma}
$$
(i)~~~~~N < \sqrt{\frac{\pi}{2g}}(1 + \alpha_N),              \eqno (C.18)
$$
$$
(ii)~~~~~N > \sqrt{\frac{\pi}{2g}}(1 - \beta_N),               \eqno (C.19)
$$
$$
(iii)~~~~~M < \frac{1}{8g}e^{\frac{4}{3}g} (1 + \alpha_M)   \eqno (C.20)
$$
and
$$
(iv)~~~~~M > \frac{1}{8g}e^{\frac{4}{3}g} (1 - \beta_M),    \eqno (C.21)
$$
where
$$
1 + \alpha_N = (1-\frac{3}{4g^\frac{1}{3}})^{-2}(1+ \frac{5}{48g}) +
(1+e^{-\frac{9}{2}})\sqrt{ \frac{2}{\pi}}~\frac{1}{ g^{1/6}} e^{-3g^{\frac{1}{3}}} +
\frac{1}{2\sqrt{2\pi g}}e^{-2g}                                             \eqno (C.22)
$$
$$
1 - \beta_N = 1 -\frac{1}{\sqrt{2\pi g}}e^{-2g}            \eqno (C.23)
$$
$$
1 + \alpha_M = ( 1-e^{-2\sqrt{g}})e^{\frac{2}{3\sqrt{g}}}(1+\frac{1}{\sqrt{g}})^2
+ 6(e^{-\frac{4}{3}\sqrt{g}}-e^{-\frac{4}{3}g})               \eqno (C.24)
$$
$$
1 - \beta_M = 1 - e^{-2g}                                   \eqno (C.25)
$$
so that for $g$ sufficiently large, $\alpha_N$, $\beta_N$,  $\alpha_M$ and
$\beta_M$ are all small, with
$$
\alpha_N = O(g^{-\frac{1}{3}}),~~~~~~~~  \beta_N = O(\frac{1}{\sqrt{g}}e^{-2g}),\eqno (C.26)
$$
$$
\alpha_M = O(g^{-\frac{1}{2}})  ~~~{\sf  and}~~~ \beta_M = O(e^{-2g})       \eqno (C.27)
$$

\noindent
\underline{Proof of the lemma}

The lower bound for $N$ can be most easily derived by setting $x=1 \mp \xi$, as in
(C.14), and similarly observing that $N=\int\limits^{\infty}_0 \phi^2_+ dx$ satisfies
$$
N > \int\limits^2_0 \frac{4}{(1+x)^2} e^{-2g S_0(x)}dx >
\int\limits^1_0 2e^{-2g \xi^2}d\xi
> \sqrt{\frac{\pi}{2g}}(1 - \frac{1}{ \sqrt{2 \pi g}}~e^{-2g})     \eqno (C.28)
$$
which gives (C.19) and (C.23).

For the upper bound of $N$, we decompose $N$ into a sum of three terms:
$$
N = N_1 + N_2 + N_3,                                \eqno (C.29)
$$
where
$$
N_1 =  \int\limits^{\infty}_2 \phi^2_+(x) dx,        \eqno (C.30)
$$
$$
N_2 =  \int\limits^{1+l}_{1-l} \phi^2_+(x) dx        \eqno (C.31)
$$
and
$$
N_3 = ( \int\limits^{1-l}_0 dx + \int\limits^2_{1+l} dx )  \phi^2_+(x).   \eqno (C.32)
$$
Note that, setting $x=1+\xi$, we obtain
$$
N_1 = \int\limits^{\infty}_1 (\frac{2}{2+\xi})^2 e^{-2g \xi^2-\frac{2}{3}g \xi^3} d\xi
 < \int\limits^{\infty}_1  e^{-2g \xi^2}d\xi < \frac{1}{4g} e^{-2g}.    \eqno (C.33)
$$

For $N_2$, the integral can be written as $ \int\limits^l_0 (\phi^2_+(x_+) + \phi^2_+(x_-)
)d\xi$ with $x_{\pm}=1 \mp \xi$, as in (C.14). Since
$$
N_2 =  \int\limits^l_0 (\frac{4}{(2-\xi)^2}e^{\frac{2}{3}g \xi^3} +
 \frac{4}{(2+\xi)^2}e^{-\frac{2}{3}g \xi^3}) e^{-2g \xi^2} d\xi
 <  \int\limits^l_0 \frac{8}{(2-\xi)^2} \cosh (\frac{2}{3}g \xi^3) e^{-2g \xi^2}
 d\xi,                                                          \eqno (C.34)
$$
by choosing $l$ to satisfy
$$
\frac{2}{3}g l^3=(\frac{3}{2})^2,                                       \eqno (C.35)
$$
we have for
$$
\xi \le l = \frac{3}{2}~\frac{1}{g^{1/3}}, \eqno (C.36)
$$
$$
(1-\frac{\xi}{2})^{-2} < (1-\frac{3}{4}~\frac{1}{g^{1/3}})^{-2},             \eqno (C.37)
$$
$$
\cosh(\frac{2}{3} g\xi^3) < 1+ ( \frac{2}{3} g\xi^3)^2,                       \eqno (C.38)
$$
and therefore
$$
N_2 < 2(1- \frac{3}{4g^{1/3}})^{-2} \int\limits_0^l e^{-2g\xi^2} (1+ \frac{4}{9}g^2\xi^6)d\xi
\nonumber
$$
$$
~~~~<  2(1- \frac{3}{4g^{1/3}})^{-2} \int\limits_0^{\infty} e^{-2g\xi^2}
 (1+ \frac{4}{9}g^2\xi^6)d\xi
= (1- \frac{3}{4g^{1/3}})^{-2} \sqrt{\frac{\pi}{2g}}(1+ \frac{5}{48g}).      \eqno (C.39)
$$

Likewise, from (C.32)
$$
N_3 < 8  \int\limits_l^1 e^{-2g\xi^2}\cosh ( \frac{2}{3} g\xi^3)
d\xi\nonumber
$$
$$
= 4  \int\limits_l^1 e^{-2g\xi^2+  \frac{2}{3} g\xi^3}
(1+ e^{-\frac{4}{3}g\xi^3})d\xi \nonumber
$$
$$
< 4 (1+ e^{-\frac{4}{3}gl^3})\int\limits_l^1
e^{-\frac{4}{3}g\xi^2}d\xi \nonumber
$$
$$
< 4 (1+ e^{-\frac{4}{3}gl^3})\int\limits_l^{\infty}
e^{-\frac{4}{3}g\xi^2}d\xi \nonumber
$$
$$
<(1+ e^{-\frac{4}{3}gl^3})\frac{3}{2g l}~e^{-\frac{4}{3}gl^2}
<(1+e^{-\frac{9}{2}})~\frac{1}{ g^{2/3}} e^{-3g^{\frac{1}{3}}},
 \eqno (C.40)
$$
in which, because of (C.35), $e^{-\frac{4}{3}gl^2} = e^{-3g^{1/3}}$. Combining (C.33),
(C.39) and (C.40), we derive the upper bound for $N$, given by
(C.18) and (C.22).

Next, we write the integral for $M$, defined by (C.5), as
$$
M = \big(\int\limits_0^a dx + \int\limits_a^1 dx\big)( \frac{1+x}{2})^2
e^{-2gx + \frac{2}{3}gx^3 + \frac{4}{3}g}. \eqno (C.41)
$$
Choose
$$
a = 1/\sqrt{g} \eqno (C.42)
$$
so that $ga^3 = a = 1/\sqrt{g}$ and
$$
M < ( \frac{1+a}{2})^2
e^{\frac{2}{3}ga^3+\frac{4}{3}g}\int\limits_0^ae^{-2gx}dx
+\int\limits_a^1 ( \frac{1+x}{2})^2
e^{-\frac{4}{3}g x+\frac{4}{3}g}dx\nonumber
$$
$$
~~~< ( \frac{1+a}{2})^2
e^{\frac{2}{3}ga^3+\frac{4}{3}g}\frac{1}{2g}(1-e^{-2ga})
+\int\limits_a^1 e^{-\frac{4}{3}g x+\frac{4}{3}g}dx\nonumber
$$
$$
~~~= \frac{1}{8g}~e^{\frac{4}{3}g}\big\{
(1-e^{-2\sqrt{g}})~e^{\frac{2}{3\sqrt{g}}}
(1+\frac{1}{\sqrt{g}})^2 + 6 (e^{-\frac{4}{3}\sqrt{g}}-e^{-\frac{4}{3}g})\big\},\eqno (C.43)
$$
which gives (C.20) and (C.24).

For the lower bound, we have from (C.41)
$$
M > \int\limits_0^1\frac{1}{4}~ e^{-2gx + \frac{4}{3}g} dx
= \frac{1}{8g}~e^{ \frac{4}{3}g}(1-e^{-2g}),  \eqno (C.44)
$$
which gives (C.21) and (C.25) and completes the proof for the
lemma.

From (C.17)-(C.18) and (C.20), we derive the lower bound for
$\gamma$, determined by
$$
\gamma^{-1} < \frac{1}{4g}\sqrt{\frac{\pi}{2g}}~e^{\frac{4}{3}g}~\big[~(1+\alpha_M)
(1+\alpha_N)+ 4g\sqrt{\frac{2g}{\pi}}~e^{-\frac{4}{3}g}~ J_+~\big], \eqno (C.45)
$$
which leads to (C.4).
\\

\noindent

C.2~~~Bounds for $\psi_-$, $\psi_-'$ and $\chi'$

(i) Let
$$
{\cal M}(x) \equiv \int\limits_x^1 \phi_+^{-2}(y)dy  \eqno(C.46)
$$
and write
$$
\phi_+(x) = e^{-gS(x)} \eqno(C.47)
$$
with
$$
S(x) \equiv S_0(x) + \frac{1}{g}\ln\frac{1+x}{2} =
\frac{1}{3}(x-1)^2(x+2) +\frac{1}{g}\ln\frac{1+x}{2}.   \eqno(C.48)
$$
We shall first establish: for
$$
0<x<1-g^{-1},\eqno(C.49)
$$
$$
{\cal M}(x) < \frac{1-x}{2gS(x)}\phi_+^{-2}(x).  \eqno(C.50)
$$

\noindent
\underline{Proof}~~~~For $0<x<y<1$, define
$$
\overline{S}(y) \equiv S(x)\frac{1-y}{1-x} ~.   \eqno(C.51)
$$
One can readily show that
$$
S(y) < \overline{S}(y).   \nonumber
$$
Therefore,
$$
{\cal M}(x) < \int\limits_x^1 e^{2g\overline{S}(y)} dy
< \frac{1-x}{2gS(x)}(e^{2gS(x)}-1) < \frac{1-x}{2gS(x)}~e^{2gS(x)},\nonumber
$$
which gives (C.50). The restriction of $x<1-g^{-1}$ in (C.49) is to avoid the
uninteresting complication of $S(x)$ having a zero at $x \cong 1-\frac{1}{2}g^{-1}$.
For $1>x>1-g^{-1}$, since $S(x)<g^{-2}$, we have $\phi_+^{-2}<e^{2/g}$, and
therefore on account of (C.46)
$$
{\cal M}(x)<\frac{1}{g}~e^{2/g}=\frac{1}{g}~(1+O(\frac{1}{g})~).    \eqno(C.52)
$$

(ii)~~~Next, we turn to $\psi_-(x)$. From (\ref{e5.5}),
$$
\frac{\psi_-}{\psi_+} = 2\gamma \int\limits_x^{\infty}\psi_+^{-2}(y)dy
\int\limits_y^{\infty}\psi_+^2(z)dz
< 2\gamma \int\limits_x^{\infty}\phi_+^{-2}(y)dy
\int\limits_y^{\infty}\phi_+^2(z)dz,
$$
since $\psi_+ = f_+ \phi_+$ and $f_+(z) < f_+(y)$ for $z>y$. Thus, for $x>1$,
$$
\frac{\psi_-}{\psi_+} < \gamma J_+,    \eqno(C.53)
$$
where $\gamma \cong 4g\sqrt{\frac{2g}{\pi}}~e^{-\frac{4}{3}g}$ and
$J_+ < \frac{1}{2g}\ln(e+2e\sqrt{2\pi g})$ are bounded by (C.1)-(C.2) and (\ref{e4.29}).
For $x<1$
$$
\frac{\psi_-(x)}{\psi_+(x)}<\frac{\psi_-(1)}{\psi_+(1)}
+ 2\gamma \int\limits_x^1\phi_+^{-2}(y)dy
\int\limits_y^{\infty}\phi_+^2(z)dz < \gamma (J_+ + 2N{\cal M}(x)),  \eqno(C.54)
$$
where $N=\int\limits_0^{\infty}\phi_+^2(z)dz $ is bounded by (C.18)-(C.19) and
${\cal M}(x)$ by (C.50) and (C.52). Hence, for $1>x>0$,
$$
\psi_-(x)\psi_+(x)< \gamma \psi_+^2(x)(J_+ + 2N {\cal M}(x)) = O(e^{-\frac{4}{3}g}).
           \eqno(C.55)
$$
It follows then, for all $x \ge 0$,
$$
\psi_-(x)\psi_+(x) \le O(e^{-\frac{4}{3}g}).                    \eqno(C.56)
$$

At $x=0$, on account of (\ref{e5.6}),
$$
\psi_-(0)\psi_+(0)=\psi_+^2(0)=4f_+^2(0)e^{-\frac{4}{3}g}.
$$
Since $\psi_+$ increases from $O(e^{-\frac{2}{3}g})$ at $x=0$ to $O(1)$ at $x=1$,
$\psi_-$ decreases from  $O(e^{-\frac{2}{3}g})$ at $x=0$ to $O(e^{-\frac{4}{3}g})$
at $x=1$.

Recall that, from (2.6)-(2.7),
$$
\phi_-(x)\phi_+(x)=\big( \frac{2}{1+x}\big)^2 e^{-\frac{4}{3}g},  \eqno(C.57)
$$
which, at $x=0$, is
$$
\phi_-(0)\phi_+(0)= 4 e^{-\frac{4}{3}g}.    \eqno(C.58)
$$
Its derivative is
$$
\phi_+\phi_-' + \phi_-\phi_+' = -\frac{2}{1+x}\phi_+\phi_-;    \nonumber
$$
at $x=0$, the two terms on the left hand side are
$$
\phi_-(0)\phi_+'(0)= 4 (g-1) e^{-\frac{4}{3}g}        \nonumber
$$
and
$$
\phi_+(0)\phi_-'(0)= -4 (g+1) e^{-\frac{4}{3}g}.       \eqno(C.59)
$$
Furthermore, for $0<x<1$
$$
\phi_-'(x) = \Big(-g(1-x^2) - \frac{1}{1+x}\Big) \phi_- <0, \nonumber
$$
and therefore the derivative of the difference $\phi_+ - \phi_-$ satisfies
$$
\phi_+' - \phi_-' > \phi_+' = -g S' \phi_+. \eqno(C.60)
$$
The above expressions (C.55)-(C.58) describe the similarity between $\phi_-(x)\phi_+(x)$
and  $\psi_-(x)\psi_+(x)$.
As we shall see, the derivatives $\psi_+'$,  $\psi_-'$ and  $\chi'=\psi_+' - \psi_-'$
also satisfy inequalities similar to (C.59)-(C.60).

(iii) ~~In this subsection, we shall establish that, for
$ 0<x< 1-\sqrt{\frac{\ln g}{g}}$,
$$
\chi'(x) > \psi_+'(x) \Big(1-O(e^{-\frac{4}{3}g})\Big).   \eqno(C.61)
$$

\noindent
\underline{Proof}~~~From (\ref{e4.43}), $\psi_+ = f_+ \phi_+ > \phi_+$.
By using (\ref{e5.5}) we see that
$$
\psi_+\psi_-' = \psi_-\psi_+' - 2 \gamma \int\limits_x^{\infty} \psi_+^2(z) dz
< \psi_-\psi_+' - 2 \gamma \int\limits_x^{\infty} \phi_+^2(z) dz,   \eqno(C.62)
$$
which, at $x=0$, becomes
$$
\psi_+(0)\psi_-'(0) < \psi_-(0)\psi_+'(0) - 2 \gamma N,    \eqno(C.63)
$$
where, on account of (C.19) and (C.23),
$$
N = \int\limits_0^{\infty} \phi_+^2(x)dx > \sqrt{\frac{\pi}{2g}}
(1-\frac{1}{\sqrt{ 2\pi g}}~e^{-2g}).    \eqno(C.64)
$$

Because of (\ref{e4.42}) and (C.47)-(C.48),
$$
\frac{\psi_+'}{\psi_+} = \frac{f_+'}{f_+} + \frac{\phi_+'}{\phi_+} <
\frac{\phi_+'}{\phi_+} = -g S' = g(1-x^2) - \frac{1}{1+x}.     \eqno(C.65)
$$
Multiplying (C.56),
$$
\psi_- \psi_+ < \gamma \psi_+^2(J_+ + 2 N {\cal M}),        \eqno(C.66)
$$
by $\psi_+'/\psi_+$ and using  (C.65), we find
$$
\psi_- \psi_+' < \gamma J_+ \psi_+ \psi_+' + 2 \gamma N {\cal M} \psi_+ \psi_+'
< \gamma J_+ \psi_+ \psi_+' - 2 \gamma N {\cal M} \psi_+^2 g S'.
$$
Substituting this expression into (C.62), we derive
$$
\psi_+ \psi_-'< \gamma J_+ \psi_+ \psi_+' -
 2 \gamma({\cal N} + g N {\cal M} \psi_+^2 S'),      \eqno(C.67)
$$
where
$$
{\cal N}(x) \equiv \int\limits_x^{\infty} \phi_+^2(z)dz.    \eqno(C.68)
$$

At $x=0$,
$$
{\cal M}(0) = M                                                                  \eqno(C.69)
$$
and
$$
{\cal N}(0) = N.                                                                 \eqno(C.70)
$$
where $M$ and $N$ are given by (C.5)-(C.6).

On account of (C.65) and (\ref{e4.44}),
$$
\psi_+(0)\psi_+'(0)<\psi_+^2(0)(g-1) = f_+^2(0) 4(g-1)~e^{-\frac{4}{3}g}
<(1-L)^{-2}4(g-1)~e^{-\frac{4}{3}g} \cong 4g~e^{-\frac{4}{3}g},                  \eqno(C.71)
$$
where $L=O(\frac{1}{g}\ln\sqrt{g})$. From (C.1)-(C.2), we see that
$\gamma = O(\epsilon)$, where $\epsilon = e^{-\frac{4}{3}g}$ as before. Therefore,
$$
\gamma J_+ \psi_+^2(0) = O(\epsilon^2)~~~{\sf and}~~~
\gamma J_+ \psi_+(0)\psi_+'(0) = O(\epsilon^2)                                   \eqno(C.72)
$$
can both be neglected in (C.66) and (C.67). Because
$$
2 \gamma {\cal N}(0) = 2 \gamma N > 2\cdot 4g\sqrt{\frac{2g}{\pi}}(1-\beta_{\gamma})
\sqrt{\frac{\pi}{2g}}(1-\beta_{\cal N})~e^{-\frac{4}{3}g} \cong 8g~e^{-\frac{4}{3}g}
                                                                                 \eqno(C.73)
$$
and
$$
-g{\cal M}(0) \psi_+^2(0) S'(0) < \frac{1}{8g}~e^{\frac{4}{3}g}
(1+\alpha_M)(1-L)^{-2}4~e^{-\frac{4}{3}g}(g-1) \cong \frac{1}{2}~,               \eqno(C.74)
$$
it follows then
$$
\psi_-(0)\psi_+'(0) < \gamma J_+ \psi_+(0)\psi_+'(0) -
2 \gamma N M g \psi_+^2(0)S'(0) \cong 4g~e^{-\frac{4}{3}g}                       \eqno(C.75)
$$
and
$$
\psi_+(0)\psi_-'(0) < \gamma J_+ \psi_+(0)\psi_+'(0) -
2 \gamma N (1+ g M \psi_+^2(0)S'(0))\nonumber
$$
$$~~~~~~~~~~~ \cong -8g~e^{-\frac{4}{3}g}(1-\frac{1}{2})
=-4g~e^{-\frac{4}{3}g}                                                           \eqno(C.76)
$$
similar to (C.59).

Next, consider the region
$$
0<x<1-\sqrt{\frac{\ln g}{g}}                                                     \eqno(C.77)
$$
so that the integral (C.68) for ${\cal N}(x)$ already contains the Gaussian peak of
$\phi_+^2(z)$ in its integrand. Hence, in the region (C.77), by
following the same reasoning given in (C.28) and writing
$x=1-\xi$, we find
$$
{\cal N}(x)=\int\limits_{1-\xi}^{\infty} \phi_+^2(y)dy >
\int\limits_{1-\xi}^{1+\xi} \phi_+^2(y)dy > 2\int\limits_0^{\xi}
e^{-2gy^2}dy >\sqrt{ \frac{\pi}{2g}}(1- \frac{1}{\sqrt{2\pi g }}~\frac{1}{\xi}~
e^{-2g \xi^2})~.\nonumber
$$
For $x$ within the range (C.77), the smallest value of $\xi$ is
$\sqrt{\frac{\ln g}{g}}$. Therefore
$$
{\cal N}(x) > \sqrt{\frac{\pi}{2g}}(1-\beta_{{\cal N}})= \sqrt{\frac{\pi}{2g}}
\Big(1-\frac{1}{g^2} \frac{1}{\sqrt{2\pi \ln g}}\Big).                            \eqno(C.78)
$$

  In what follows, we shall first show
$$
C(x) \equiv \psi_+(x)\psi_-'(x)- \gamma J_+ \psi_+(x)\psi_+'(x)                  \eqno(C.79)
$$
is negative in the region (C.77).

  From (C.67) by using (C.78) for ${\cal N}(x)$, (C.50) for ${\cal M}(x)$ and (\ref{e4.9}) for
$\psi_+(x)$, we see that
$$
C(x) < -2\gamma N \Big[(1-\beta_{\cal N})(1+\alpha_N)^{-1}
+ \frac{g(1-x)S'(x)}{2gS(x)}~f_+^2(x)\Big].
                                                                                 \eqno(C.80)
$$
For $g$ sufficiently large we can
approximate $1-\beta_{{\cal N}}~$, $1+\alpha_N$ and $f_+^2(x)$ by $1$. Hence,
$$
C(x) < \overline{C}(x) \cong 2\gamma N \Big(-1-\frac{g(1-x)S'(x)}{2gS(x)}\Big),  \eqno(C.81)
$$
in which
$$
-\frac{g(1-x) S'(x)}{2gS(x)} \cong -\frac{1}{2}~\frac{(1-x)}{S_0(x)}S_0'(x)
=\frac{(1-x)(1-x^2)}{2(1-x)^2\frac{1}{3}(2+x)} = \frac{3(1+x)}{2(2+x)}<1.        \eqno(C.82)
$$
Therefore for $g$ sufficiently large, and within the region (C.77),
$$
C(x) = \psi_+(x)\psi_-'(x) -\gamma J_+  \psi_+(x)\psi_+'(x) < 0.                 \eqno(C.83)
$$

We now turn to the derivative of
$$
\chi(x) = \psi_+(x) - \psi_-(x).        \nonumber
$$
Since
$$
\psi_+(x)\chi'(x) = \psi_+(x)\psi_+'(x)-\psi_+(x)\psi_-'(x)
=\psi_+(x)\psi_+'(x)(1-\gamma J_+) -C(x).                                        \eqno(C.84)
$$
On account of (C.83),
$$
\psi_+(x)\chi'(x) > \psi_+(x)\psi_+'(x)(1-\gamma J_+)
=\psi_+(x)\psi_+'(x)\big(1-O(\epsilon)\big),                                     \eqno(C.85)
$$
which completes the proof of (C.61). As we shall see, this result greatly simplifies
the derivation of the bound for ${\cal I}$.
\\

\noindent

C.3~~~~Bound for ${\cal I}$

We begin with (\ref{e5.58}),
$$
{\cal I} = 2\int\limits_0^1 \chi^{-2}(y)dy
\int \limits_0^y \chi(z) \psi_+(z) dz.          \nonumber
$$
As we shall establish,
$$
{\cal I} < 2 \Big( \frac{3\pi}{g^3}\Big)^{1/4} (1+ \alpha_{{\cal I}}),         \eqno(C.86)
$$
where $1+\alpha_{{\cal I}} = 1+O(g^{-\frac{1}{4}})$ is given by
$$
1+\alpha_{\cal I} = \frac{1}{2}(1-L)^{-2}
+ \frac{1}{2}[1- \frac{1}{2}(3\pi g)^{-\frac{1}{4}}]\cdot
\Big\{1-\frac{1}{4}(3\pi g)^{-\frac{1}{4}}
-\frac{1}{2} (3\pi)^{\frac{1}{4}} g^{-\frac{3}{4}}
[1- \frac{1}{4}(3\pi g)^{-\frac{1}{4}}]^{-1}
$$
$$
-\frac{3}{2}(3\pi)^{\frac{3}{4}}
g^{-\frac{5}{4}}(1-L)^{-1}\Big\}^{-1}                                         \eqno(C.87)
$$
and $L=O(\frac{1}{g} \ln \sqrt{g})~$ is bounded by (4.45).

\noindent
\underline{Proof}~~~~Divide the integration range into two regions: 1. $c<y<1$ and
 2. $0<y<c$. Correspondingly,
$$
{\cal I} = {\cal I}_1 + {\cal I}_2,                                              \eqno(C.88)
$$
where
$$
{\cal I}_1 = 2\int\limits_c^1 \chi^{-2}(y)dy
\int \limits_0^y \chi(z) \psi_+(z) dz                                            \eqno(C.89)
$$
and
$$
{\cal I}_2 = 2\int\limits_0^c \chi^{-2}(y)dy
\int \limits_0^y \chi(z) \psi_+(z) dz.                                           \eqno(C.90)
$$
choose
$$
c=1- \frac{1}{2}(3\pi g)^{-\frac{1}{4}}~.                                       \eqno(C.91)
$$
Hence, in region 1,
$$
\psi_+(y)>\psi_+(c) = f_+(c) \frac{2}{1+c}~e^{-g( \frac{1}{4}(\frac{1}{3\pi g})^{1/2}
\frac{1}{3}(2+c)}~=~O\Big(e^{- \frac{1}{4}(\frac{g}{3\pi})^{1/2}}\Big);    \nonumber
$$
whereas, according to (C.56) and for $g$ sufficiently large
$$
\frac{\psi_-(y)}{\psi_+(y)} = \frac{O(e^{-\frac{4}{3}g})}{\psi_+^2(y)}
\le O(e^{- \frac{4}{3}g})                                        \eqno(C.92)
$$
and therefore can be neglected. Correspondingly
$$
{\cal I}_1 \cong 2\int\limits_c^1 \psi_+^{-2}(y)dy
\int \limits_0^y \psi_+^2(z) dz.                                                 \eqno(C.93)
$$

In region 2, on account of (C.85)
$$
\chi'(z) > \psi_+'(z) (1-O(\epsilon)).   \nonumber
$$
Neglecting $O(\epsilon)$ and using $\chi^2(y) = 2 \int\limits_0^y \chi(z) \chi'(z) dz$
we have
$$
{\cal I}_2 = 2\int\limits_0^c dy \frac{\int \limits_0^y \chi(z)\psi_+(z) dz}
{2 \int \limits_0^y \chi(z) \chi'(z) dz}~<~
2 \int \limits_0^c dy \frac{\int \limits_0^y \chi(z)\psi_+(z) dz}
{2 \int \limits_0^y \chi(z) \psi_+'(z) dz} < c/\lambda,                        \eqno(C.94)
$$
where
$$
\lambda = \min \frac{ \psi_+'(z)}{ \psi_+(z)}~~~~~~~~~{\sf for}~~~0<z<c.         \eqno(C.95)
$$

To give an explicit form for both ${\cal I}_1$ and ${\cal I}_2$, we assume first $g$ very
large so that only terms with the dominant power in $g$ will be kept. Thus
$$
{\cal I}_1 \cong 2\int\limits_0^1 \phi_+^{-2}(y)dy
\int \limits_0^y \phi_+^2(z) dz.      \nonumber
$$
From (A.62),
$$
\phi_+^{-2}(y)\phi_+^2(z)=( \frac{1+y}{1+z})^2~ e^{-2g(S_0(y)-S_0(z))}
<4~e^{-\frac{4}{3}g(y-z)^2}~,                                                    \eqno(C.96)
$$
which leads to
$$
{\cal I}_1 < 2\sqrt{\frac{3\pi}{g}} (1-c).                                       \eqno(C.97)
$$

Likewise, in region 2,
$$
\frac{\psi_+'(z)}{\psi_+(z)} \cong \frac{\phi_+'(z)}{\phi_+(z)} \cong -gS_0'(z) > g(1-c^2).
     \nonumber
$$
Thus,
$$
{\cal I}_2 < \frac{c}{g(1-c^2)} \cong \frac{1}{2g(1-c)}.                         \eqno(C.98)
$$
The optimal $c$ is $1-\frac{1}{2}(3\pi g)^{-\frac{1}{4}}$, which yields the leading form of
${\cal I}$ given by (C.86). The inclusion of non-leading order terms is tedious,
as will be given below.

For our purpose, we only need to establish a finite upper bound
for ${\cal I}$. For $g$ large, it can be readily verified that the
neglect of $O(\epsilon)$ terms is fully justified. Thus, from
(C.93) and (C.97), we obtain
$$
{\cal I}_1 < 2\sqrt{\frac{3\pi}{g}}(1-c)f_+^2(0) <
2\sqrt{\frac{3\pi}{g}}(1-c)(1-L)^{-2}                                           \eqno(C.99)
$$
where $L=O(\frac{1}{g} \ln \sqrt{g} )$ is bounded by (4.45).

  Likewise, (C.94) and (C.95) are valid after the neglect of
 $O(\epsilon)$ terms. For ${\cal I}_2$, we observe that
$$
\frac{\psi_+'(z)}{\psi_+(z)}=\frac{f_+'(z)}{f_+(z)}+\frac{\phi_+'(z)}{\phi_+(z)}. \eqno(C.100)
$$
From (4.55) and using $f_+(0)/f_+(z)<(1-L)^{-1}$, we have
$$
\frac{f_+'(z)}{f_+(z)}>-\frac{3}{2} \sqrt{\frac{3\pi}{g}}(1-L)^{-1}~.           \eqno(C.101)
$$
In addition,
$$
\frac{\phi_+'(z)}{\phi_+(z)}=g(1-z^2)-\frac{1}{1+z}                             \eqno(C.102)
$$
which at $z=c=1-\frac{1}{2}(3\pi g)^{-\frac{1}{4}}$ gives
$$
\frac{\phi_+'(c)}{\phi_+(c)}
= g (3\pi g)^{-\frac{1}{4}}~
\Big[ 1-\frac{1}{4}(3\pi g)^{-\frac{1}{4}}-
\frac{1}{2} (3\pi)^{\frac{1}{4}} g^{-\frac{3}{4}}
\Big(1-\frac{1}{4}(3\pi g)^{-\frac{1}{4}}\Big)^{-1} \Big] \nonumber
$$
$$
= 2 g(1-c)(1+O(g^{-\frac{1}{4}}))~.                                  \eqno(C.103)
$$
Hence,
$$
{\cal I}_2 < \frac{1}{2g(1-c)}(1+\alpha_{{\cal I}_2})                  \eqno(C.104)
$$
where
$$
1+\alpha_{{\cal I}_2}=[1-\frac{1}{2}(3\pi g)^{-\frac{1}{4}}]
\Big[1-\frac{1}{4}(3\pi g)^{-\frac{1}{4}}-
\frac{1}{2} (3\pi)^{\frac{1}{4}} g^{-\frac{3}{4}}
[1-\frac{1}{4}(3\pi g)^{-\frac{1}{4}}]^{-1}
$$
$$
-\frac{3}{2}(3\pi)^{\frac{3}{4}}g^{-\frac{5}{4}}(1-L)^{-1}\Big]^{-1}~.     \eqno(C.105)
$$
Combining (C.99) with (C.104)-(C.105), we have for
$1-c=\frac{1}{2}(3\pi g)^{-\frac{1}{4}}$
$$
{\cal I} < 2(\frac{3\pi}{g^3})^{\frac{1}{4}}(1+\alpha_{\cal I} )         \eqno(C.106)
$$
where $1+\alpha_{\cal I}$ is given by (C.87).
\\

\noindent

C.4~~~~Bounds for ${\cal J}^+$ and ${\cal J}^-$

  From (\ref{e5.59}) and (\ref{e5.78}), we have
$$
{\cal J}^+ = 2\int \limits_1^{\infty} \chi^{-2}(y)dy  \int \limits_y^{\infty} \chi^2(z)dz
$$
and
$$
{\cal J}^- = 2\int \limits_0^1 \chi^{-2}(y)dy  \int \limits_0^y \chi^2(z)dz~.
$$

  For $x>1$, according to (C.53)
$$
\frac{\psi_-}{\psi_+} < \gamma J_+ = O(\epsilon)~.   \nonumber
$$
Thus, on account of (\ref{e4.29}) and neglecting $O(\epsilon)$, we have
$$
{\cal J}^+ \cong J_+ < \frac{1}{2g} \ln (e+2e\sqrt{2\pi g}~).       \eqno(C.107)
$$

  For ${\cal J}^-$, because $\chi~<~\psi_+$, we have
$$
{\cal J}^-~<~{\cal I}~<~2(\frac{3\pi}{g^3})^{\frac{1}{4}}(1+\alpha_{\cal I})~. \eqno(C.108)
$$
Combining together ${\cal J}^+$ and ${\cal J}^-$, we derive
$$
{\cal J} =  {\cal J}^+ + {\cal J}^- <  \frac{1}{2g} \ln (e+2e\sqrt{2\pi g}~)
 + 2(\frac{3 \pi}{g^3})^{\frac{1}{4}} (1 + \alpha_{{\cal I}})       \eqno(C.109)
$$
\\

\noindent
C.5~~~~Bound for $\Delta_1$

  Set $n=1$ in (\ref{e5.43}). Since $k_0=1$ and, according to (\ref{e5.10}),
  $\nu \chi = -\gamma \psi_+$, we find
$$
\Delta_1 = \gamma \frac{\int\limits_0^{\infty}\chi \psi_+ dx}
{\int\limits_0^{\infty}\chi^2 dx}
<  \gamma \frac{\int\limits_0^{\infty}\chi \psi_+ dx}
{\int\limits_l^{\infty}\chi^2 dx}                                               \eqno(C.110)
$$
where, for convenience, we choose
$$
l=1-\sqrt{\frac{\ln g}{g}}                                                      \eqno(C.111)
$$
so that (i) for $x>l$, as can be readily seen from (C.53) and (C.56),
$$
\frac{\psi_-(x)}{\psi_+(x)} \sim O(\epsilon)                                    \eqno(C.112)
$$
and (ii) $x=l$ is already outside the Gaussian peak of
$\phi_+^2(x)$, which is near $x=1$ with a width
$O(\frac{1}{\sqrt{g}})$. Using (C.112) and neglecting $O(\epsilon)$,
we can approximate the denominator in the righthand side of (C.110) as
$$
\int\limits_l^{\infty}\chi^2 dx \cong  \int\limits_l^{\infty} \psi_+^2 dx
> \int\limits_l^{\infty} \phi_+^2 dx~.                                       \eqno(C.113)
$$

  For the numerator in (C.110), we note that
$$
\chi=\psi_+ - \psi_- < \psi_+;
$$
therefore,
$$
\int\limits_0^{\infty}\chi \psi_+ dx <
\int\limits_0^{\infty}\psi_+^2 dx < f_+^2(0) \int\limits_0^{\infty}\phi_+^2 dx~.\eqno(C.114)
$$
From (C.110) we derive, after neglecting the $O(\epsilon)$
correction,
$$
\Delta_1 < \gamma f_+^2(0) \frac{\int\limits_0^{\infty} \phi_+^2 dx}
{\int\limits_l^{\infty} \phi_+^2 dx}~.                                      \eqno(C.115)
$$

From (C.6) and (C.78), we have
$$
\int\limits_0^{\infty} \phi_+^2(x) dx < \sqrt{\frac{\pi}{2g}} (1 +
\alpha_N),               \nonumber
$$
$$
\int\limits_l^{\infty} \phi_+^2(x) dx > \sqrt{\frac{\pi}{2g}} \Big(
1-\frac{1}{g^2\sqrt{2\pi \ln g}}\Big),
$$
and therefore, for large $g$, $\Delta_1 \cong \gamma$, with an upper bound given by
$$
\frac{\Delta_1}{\gamma} < (1-L)^{-2} (1+ \alpha_N)
\Big(1-\frac{1}{g^2\sqrt{2\pi \ln g}}\Big)^{-1},                          \eqno(C.116)
$$
where $L=O( \frac{\ln~\sqrt{g}}{g})$ and $\alpha_N =
O(g^{-\frac{1}{3}})$ are given by (\ref{e4.45}) and (C.8).

\section*{\bf Appendix D}
\setcounter{section}{10}
\setcounter{equation}{0}

\noindent
\underline{D.1}~~~
In this section we examine the first iterated solution $f_1$ and its associated
energy ${\cal E}_1$, given by
(\ref{e1.31}) and  (2.32). From (\ref{e1.31}), we see that the derivative
 of $f_1$ is given by
$$
f_1'(x)= 2 \phi^{-2}(x)
\int\limits_x^{\infty} \phi^2(y)(w(y)-{\cal E}_1) dy,                   \eqno(D.1)
$$
where because of (2.32),
$$
f_1'(0)=0.                                                            \eqno(D.2)
$$
The energy ${\cal E}_1$ and the functions $f_1(x)$ and $f_1'(x)$ are expressed in
terms of integrals which are well defined for any $g>0$. However, it is of interest
to expand $f_1'$ and ${\cal E}_1$ in power series of $g^{-1}$:
$$
f_1'= \sum\limits^{\infty}_1 \frac{1}{g^m} \sigma_{m+1}'=\frac{1}{g}\sigma_2'
+\frac{1}{g^2}\sigma_3'+\cdots
$$
$$
{\sf and}                                                             \eqno(D.3)
$$
$$
{\cal E}_1 = \sum\limits^{\infty}_0 \frac{1}{g^m} e_{m+1}= e_1 + \frac{1}{g} e_2
+\frac{1}{g^2} e_3+\cdots.
$$
As we shall see, these expansions give only asymptotic series.

For such power series expansion, we can neglect terms $O(e^{-\frac{4}{3}g})$ and
approximate (2.8) and  (D.1) by
$$
\phi(x) \cong \phi_+(x)                                            \eqno(D.4)
$$
and
$$
f_1'(x) \cong 2\phi_+^{-2}(x)\int\limits_x^{\infty} \phi_+^2(y)
(u(y) -  {\cal E}_1)dy,                                         \eqno(D.5)
$$
where $\phi_+$ and $u$ are given by (2.6) and (\ref{e1.11}).
Thus, the power series expansions of $f_1(x) $ and ${\cal E}_1$ are identical to
those of $f_{+,1}(x) $ and ${\cal E}_{+,1}$, introduced in (\ref{e4.15}) and
(\ref{e4.16}) when $n=1$, provided $x$ is restricted to $x \ge 0$.
We  write
$$
f_1(x) \cong f_{+,1}(x) ~~~~~~{\sf and}~~~~~{\cal E}_1 \cong {\cal E}_{+,1},
$$
where $f_{+,1}(x) $ and ${\cal E}_{+,1}$ satisfy
$$
(\frac{1}{2}\phi_+^2f_{+,1}')' = ({\cal E}_{+,1}-u) \phi_+^2~,
$$
which, on account of $\phi_+= \exp({-( g S_0 + S_1)})$, leads to
$$
g S_0'f_{+,1}'= \frac{1}{2}f_{+,1}'' - S_1'f_{+,1}'- {\cal E}_{+,1} +u.  \eqno(D.6)
$$
At $x=0$, according to (\ref{e4.19}), we have
$$
f'_{+,1}(0) = 0,                                                        \eqno(D.7)
$$
similar to (D.2).

Since $f_{+,1}(x) $ and ${\cal E}_{+,1}$ have the same formal power series expansion
as $f_1(x) $ and ${\cal E}_1$, we may write in accordance with (D.3)
$$
f_{+,1}'= \sum\limits^{\infty}_1 \frac{1}{g^m} \sigma_{m+1}'
~~~~{\sf and}~~~~
{\cal E}_{+,1} = \sum\limits^{\infty}_0 \frac{1}{g^m} e_{m+1}.               \eqno(D.8)
$$

Equating the terms proportional to $g^{-n}$ on both sides of (D.6) and recalling that\\
$u(x) = (1+x)^{-2}$, we derive
$$
S_0'\sigma_2'= \frac{1}{(1+x)^2} - e_1,                                   \eqno(D.9)
$$
$$
S_0'\sigma_3'= \frac{1}{2}\sigma_2'' - \frac{1}{1+x} \sigma_2' - e_2     \eqno(D.10)
$$
and for $m \geq 2$
$$
S_0'\sigma_{m+1}'= \frac{1}{2}\sigma_m'' - \frac{1}{1+x} \sigma_m' - e_m.  \eqno(D.11)
$$
Since
$$
S_0'= x^2-1                                                              \eqno(D.12)
$$
is zero at $x=1$, it follows from  (D.9) that
$$
e_1 = \frac{1}{4}                                                       \eqno(D.13)
$$
and
$$
\sigma_2'= - \frac{1}{2^4}(\frac{2}{1+x})^2( \frac{2}{1+x} +1).        \eqno(D.14)
$$
Likewise, from (D.10)
$$
e_2 =  \frac{9}{2^6}                                                   \eqno(D.15)
$$
and
$$
\sigma_3'= - \frac{1}{2^8}(\frac{2}{1+x})^2\Big[5 (\frac{2}{1+x})^3 +9(\frac{2}{1+x})^2
+ 9(\frac{2}{1+x}) +9 \Big].
$$
It is of interest to compare these results with the exact power series expansion of
${\cal E}$ and $f(x)$ derived in {\rm II}.
The above $e_1$, $e_2$, $\sigma_2'$ and $\sigma_3'$ are the same as the
$-E_1$, $-E_2$, $-S_2'$ and $-S_3'$ of {\rm II}, since, as can be readily
verified, they satisfy the same equations. Note that at $x=0$,
$f'_1(0)=f'_{+,1}(0)=0$, whereas $\sigma'_2(0)=-\frac{3}{4}$, $\sigma'_3(0)=-\frac{103}{64}$,
etc. Hence, the asymptotic expansion (D.3) for $f'_1$ or $f'_{+,1}$ fails completely
at $x=0$.

In general, for $m \geq 1$, we may write
$$
\sigma_{m+1}'= - \frac{1}{2^{4m}}(\frac{2}{1+x})^2\sum\limits^{2m-1}_{l=0}
\alpha_l(m)(\frac{2}{1+x})^l.                                        \eqno(D.17)
$$

\noindent
\underline{Theorem D.1}
$$
\alpha_L(m+1) =\sum\limits^{2m-1}_{l~=~{\sf max}~(0,~L-2)} (l+4)\alpha_l(m) \eqno(D.18)
$$
and
$$
e_m =  \frac{1}{2^{4m-2}}~\alpha_0(m).                                  \eqno(D.19)
$$
\underline{Proof}~~~~ From (D.17) and $S_1'=(1+x)^{-1}$, we find
$$
\frac{1}{2}\sigma_{m+1}'' - S_1' \sigma_{m+1}' = \frac{1}{2^{4m+2}}
\sum\limits^{2m-1}_{l=0}\alpha_l(m)(\frac{2}{1+x})^{l+3}(l+4).         \eqno(D.20)
$$
Since its value at $x=1$ equals $e_{m+1}$, it follows then
$$
\frac{1}{2}\sigma_{m+1}'' - S_1' \sigma_{m+1}' -e_{m+1} = \frac{1}{2^{4m+2}}\cdot
\sum\limits^{2m-1}_{l=0}\alpha_l(m)[(\frac{2}{1+x})^{l+3}-1](l+4).    \eqno(D.21)
$$
Using
$$
(\frac{2}{1+x})^{l+3}-1 = (\frac{2}{1+x}-1)[(\frac{2}{1+x})^{l+2} + (\frac{2}{1+x})^{l+1}
+ \cdots +1],                                                        \eqno(D.22)
$$
we obtain
$$
\sigma_{m+2}'= -\frac{1}{2^{4(m+1)}}(\frac{2}{1+x})^2
\sum\limits^{2m-1}_{l=0}\alpha_l(m)(l+4)[(\frac{2}{1+x})^{l+2}+(\frac{2}{1+x})^{l+1}
+\cdots +1],                                                          \eqno(D.23)
$$
and derive (D.18) and (D.19).

It is of interest to construct the following pyramid structure of $\alpha_l(m)$:
\begin{eqnarray*}
\begin{array}{ccccccl}
~~~~~~&&\alpha_1(1)&\alpha_0(1)&&&~~~m=1\\
&\alpha_3(2)&\alpha_2(2)&\alpha_1(2)&\alpha_0(2)&&~~~m=2\\
\alpha_5(3)&\alpha_4(3)&\alpha_3(3)&\alpha_2(3)&\alpha_1(3)&\alpha_0(3)&~~~m=3
~~~~~~~~~~~~~~~~~~~~~~~~~~~~~(D.24)\\
&&\cdots&\cdots&&
\end{array}
\end{eqnarray*}
By using (D.18), we see that each row $\alpha_l(m+1)$ can be obtained from
the row $\alpha_l(m)$ above. For example, using  $\alpha_1(1)=\alpha_0(1)=1$, we
have
\begin{eqnarray}
\alpha_3(2)= \alpha_1(1)\cdot (1+4) = 5, \nonumber\\
\alpha_2(2)= \alpha_1(1) \cdot 5 + \alpha_0(1) \cdot 4  = 9, \nonumber
\end{eqnarray}
etc. The resulting elements in the pyramid assume the values
\begin{eqnarray*}\label{e3.25}
\begin{array}{ccccccccl}
~~~~~~&&&1&1&&&&~~~m=1\\
&&5&9&9&9&&&~~~m=2\\
&35&89&134&170&170&170&&~~~m=3\\
315&1027&1965&2985&3835&4515&4515&4515&~~~m=4~~~~~~~~~~~~~~~~~~~~~~~~~~(D.25)\\
&&&\cdots&\cdots&&&&
\end{array}
\end{eqnarray*}
Correspondingly, (D.19) gives (D.13) and (D.15),
$$
e_3=\frac{170}{2^{10}}=\frac{85}{2^9},~~~e_4=\frac{4515}{2^{14}},~~~{\sf etc.} \eqno(D.26)
$$
We note that in this analysis $e_m$ is determined by setting the right hand side of
(D.11) to be zero at $x=1$; otherwise, the corresponding $\sigma_{m+1}'$ would
diverge at the same $x=1$. On the other hand, according to (D.1) and
(D.3), the sum $\sum\sigma_{m+1}'(x)/g^m$ at $x=1$ is
$$
f_{+,1}'(1) = 2\int\limits_1^{\infty} \phi^2_+(y)
[-{\cal E}_{+,1} +(\frac{1}{1+y})^2] dy                                  \eqno(D.27)
$$
which is finite for any given set of $e_m$ which makes the sum
$$
{\cal E}_{+,1} = \sum\limits^{\infty}_0 \frac{1}{g^m} e_{m+1}           \eqno(D.28)
$$
finite.

\noindent
\underline{D.2}~~~~
In this section, we begin with the integral expression for $f'_{+,1}(x) $:
$$
f'_{+,1}(x) = 2 \phi_+^{-2}(x) \int\limits^{\infty}_x \phi_+^2(y)
 (u(y) - {\cal E}_{+,1}) dy,                                       \eqno(D.29)
$$
where
$$
{\cal E}_{+,1} = \frac{\int\limits^{\infty}_0 \phi_+^2(x) u(x) dx}
{\int\limits^{\infty}_0 \phi_+^2(x) dx}~                           \eqno(D.30)
$$
in accordance with (\ref{e4.15}) and (\ref{e4.16}) for $n=1$. Their expansions (D.8)
lead to the equations (D.9)-(D.11) for their expansion coefficients $\sigma_m'$ and
$e_m$. Recall that in the above section, these equations are integrated from
$x = \infty$ to $x=1$. As remarked at the end of last section, we require each
integrated solution $\sigma'_{m+1}(x)$ to be well-behaved at $x=1$. This leads to
$$
e_1=\frac{1}{4},~~~e_2=\frac{9}{2^6},~~~e_3=\frac{85}{2^9},~~~{\sf etc.},\eqno(D.31)
$$
with the corresponding $\sigma'_{m+1}(x)$ given by (D.14), (D.16) and
(D.17). On the other hand, the function $f'_{+,1}(x)$, defined by the integral
(D.29), is well-behaved at $x=1$ for \underline{any} finite
${\cal E}_{+,1}$, as can be seen by the following explicit integral derived from
(D.29):
$$
f_{+,1}'(1) = 2\int\limits_1^{\infty} e^{-\frac{2}{3}g(y-1)^2(y+2)}(\frac{2}{1+y})^2
(-{\cal E}_{+,1} +(\frac{1}{1+y})^2) dy.                          \eqno(D.32)
$$
In the following, we shall show that unless the expansion
${\cal E}_{+,1}= \sum\limits_0^{\infty} \frac{1}{g^m} e_{m+1}$ has the same coefficients
$e_{m+1}$ given by (D.31), the above $f'_{+,1}(1)$ does not have an asymptotic expansion
in $g^{-1}$; instead, it would have an expansion in $g^{-\frac{1}{2}}$.

Introducing
$$
y-1=z,
$$
we first express  (D.32) as an integral over $z$:
$$
f_{+,1}'(1) = 2\int\limits_0^{\infty} dz~e^{-2gz^2}
[-\alpha(z){\cal E}_{+,1} +\beta(z)]                                 \eqno(D.33)
$$
where
$$
\alpha(z) = e^{-\frac{2}{3}gz^3}(\frac{2}{z+2})^2                   \eqno(D.34)
$$
and
$$
\beta(z) = \alpha(z) (\frac{1}{z+2})^2.                             \eqno(D.35)
$$
Next, we decompose $\alpha(z)$ and $\beta(z)$ as sums of even and odd functions of $z$:
$$
\alpha(z) = \alpha_+(z)+\alpha_-(z)                                 \eqno(D.36)
$$
and
$$
\beta(z) = \beta_+(z)+\beta_-(z),                                   \eqno(D.37)
$$
where
$$
\alpha_{\pm}(z) = \frac{1}{2}[\alpha(z)\pm \alpha(-z)]              \eqno(D.38)
$$
and
$$
\beta_{\pm}(z) = \frac{1}{2}[\beta(z)\pm \beta(-z)].                \eqno(D.39)
$$

We then expand $ \alpha_+(z)$ and $\beta_+(z)$ as power series of $z^2$. The results are
\begin{eqnarray*}
\alpha_+(z) &=& 1+\frac{3}{4}z^2 + (\frac{2}{3} + \frac{5}{2^4}~\frac{1}{g})g z^4
         +( \frac{2}{9} + \frac{1}{3}~\frac{1}{g} + O(\frac{1}{g^2}))g^2
         z^6\\
        && +( \frac{1}{6} +O( \frac{1}{g})) g^2 z^8 +( \frac{4}{3^4} + O(
        \frac{1}{g}))g^3z^{10}
        + ( \frac{2}{3^5} +O( \frac{1}{g})) g^4 z^{12} + \cdots
        ~~~~~~~~(D.40)
\end{eqnarray*}
and
\begin{eqnarray*}
 \beta_+(z) &=& \frac{1}{4}\Big\{1+\frac{5}{2}z^2 +
           (\frac{4}{3} + \frac{35}{2^4}~\frac{1}{g})g z^4
         +( \frac{2}{9} + \frac{5}{3}~\frac{1}{g} + O(\frac{1}{g^2}))g^2
         z^6\\
         && +( \frac{5}{3^2} +O( \frac{1}{g})) g^2 z^8 +( \frac{8}{3^4} + O(
         \frac{1}{g}))g^3z^{10}
          + ( \frac{2}{3^5} +O( \frac{1}{g})) g^4 z^{12} + \cdots \Big\}~.
        ~~~~~(D.41)
\end{eqnarray*}
Correspondingly, the integrals
$$
\mu_+ \equiv \int\limits_0^{\infty} dz~e^{-2gz^2}\alpha_+(z)        \eqno(D.42)
$$
and
$$
\nu_+ \equiv \int\limits_0^{\infty} dz~e^{-2gz^2}\beta_+(z)         \eqno(D.43)
$$
are given by
$$
\mu_+ =\frac{1}{2}\sqrt{\frac{\pi}{2g}}\Big[ 1+\frac{35}{3\cdot 2^5} ~\frac{1}{g} +
     (\frac{105}{2^9} + \frac{7^2\cdot 5^2}{3^2 \cdot 2^{11}})\frac{1}{g^2} +
      O(\frac{1}{g^3})\Big]                                         \eqno(D.44)
$$
and
$$
\nu_+ =\frac{1}{8}\sqrt{\frac{\pi}{2g}}\Big[ 1+\frac{89}{3\cdot 2^5} ~\frac{1}{g} +
     (\frac{395}{3 \cdot 2^7} +
     \frac{59 \cdot 9 \cdot 7 \cdot 5 \cdot 3}{3^5 \cdot 2^{11}})\frac{1}{g^2} +
      O(\frac{1}{g^3})\Big].                                       \eqno(D.45)
$$

We now introduce a new criterion, requiring the coefficient of each power
$g^{-(n+ \frac{1}{2})}$ in
$$
\mu_+\cdot \sum\limits_0^{\infty} \frac{1}{g^m} e_{m+1} + \nu_+    \eqno(D.46)
$$
to be zero. This leads to the same set (D.31) derived before.
The resulting power series expansion of (D.33) now contains only integer
power of $g^{-1}$.

Lastly, we observe that ${\cal E}_{+,1}$ (including $O(e^{-\frac{4}{3}g})$) is given by
(D.30).
To derive its asymptotic expansion in $g^{-1}$, we can neglect
$O(e^{-\frac{4}{3}g})$ and write
$$
\phi_+^2(x) = e^{-2g(x-1)^2} \alpha                                 \eqno(D.47)
$$
with
$$
\alpha = e^{-\frac{2}{3}g(x-1)^3}(\frac{2}{1+x})^2.
$$
In terms of $x-1=z$, (D.30) becomes
$$
{\cal E}_{+,1} = \int\limits_{-1}^{\infty}e^{-2gz^2} \beta(z)dz \bigg/
\int\limits_{-1}^{\infty}e^{-2gz^2}\alpha(z)dz~,                    \eqno(D.48)
$$
where $\alpha(z)$ and $\beta(z)$ are given by (D.34) and (D.35). In the decomposition
(D.36) and (D.37), the odd functions $\alpha_-(z)$ and $\beta_-(z)$ contribute only
$O(e^{-\frac{4}{3}g})$ to ${\cal E}_{+,1}$, which do not appear in the asymptotic
expansion of ${\cal E}_{+,1}$. The $\sqrt{\pi/2g}$ factor in (D.44) and (D.45) cancel
each other in the quotient (D.48). The result is the same asymptotic expansion given
by (D.23)-(D.26), as expected.

In {\rm II} we compute the coefficients $E_m$ in the asymptotic expansion of $E_{ev}$,
given by (\ref{e1.2}):
$$
E_{ev}=g+\sum\limits_0^{\infty} \frac{1}{g^m} E_{m+1}             \eqno(D.49)
$$
with
$$
E_1=-\frac{1}{4},~~~E_2=-\frac{9}{2^6},~~~E_3=-\frac{89}{2^9},~~~{\sf etc.}\eqno(D.50)
$$

In Table 1, we list the ratios of $|E_m|/e_m$ for several selected $m$ from $1$ to
$100$.

\newpage

\vspace{1cm}

\begin{center}
{\bf Table 1}
\end{center}

\begin{tabular}{cccc}
~~~$m$~~~&~~~~~~~~~$|E_m|/e_m$~~~~~~~~~&~~~$m$~~~&~~~~~~~~~$|E_m|/e_m$~~~~~~~~~\\
&&&\\
    1~   &             1~~~~~          &  ~40    &          1.5712 \\
    2~   &             1~~~~~          &  ~50    &          1.5692 \\
    3~   &             1.0471          &  ~60    &          1.5668 \\
    4~   &             1.1103          &  ~70    &          1.5643 \\
    10   &             1.4357          &  ~80    &          1.5620 \\
    20   &             1.5589          &  ~90    &          1.5599 \\
    30   &             1.5708          &  100    &          1.5579
\end{tabular}

\vspace{.5cm}

For $m>>1$, the $E_m$ is [10]
$$
E_m \sim -\frac{6}{\pi}(\frac{3}{8})^m m!                            \eqno(D.51)
$$
and, from Table 1,
$$
e_m \approx \frac{4}{\pi}(\frac{3}{8})^m m!~.                        \eqno(D.52)
$$

\end{document}